\documentclass[%
 aip,
 pop,
 amsmath,amssymb,
 reprint,%
 floatfix,%
]{revtex4-1}

\usepackage[T1]{fontenc}
\usepackage[utf8]{inputenc}
\usepackage{mathptmx}
\usepackage{bm}% bold math
\usepackage{graphicx}% Include figure files
\usepackage{etoolbox}
\usepackage{amsthm,booktabs,array,mathrsfs}
\usepackage[hidelinks]{hyperref}
\hypersetup{pdftitle={Self-consistent mixed thermal-runaway electron equilibria},pdfauthor={M. Keu, M. Hole and A. Malaquias}}

\makeatletter
\def\@email#1#2{%
 \endgroup
 \patchcmd{\titleblock@produce}
  {\frontmatter@RRAPformat}
  {\frontmatter@RRAPformat{\produce@RRAP{*#1\href{mailto:#2}{#2}}}\frontmatter@RRAPformat}
  {}{}
}%
\makeatother

\newcommand{\RE}{\mathrm{RE}}
\newcommand{\thm}{\mathrm{th}}
\newcommand{\dd}{\,\mathrm{d}}
\newcommand{\grad}{\nabla}
\newcommand{\Dstar}{\Delta^*}
\newcommand{\ephi}{\bm e_\varphi}
\newcommand{\norm}[1]{\left\lVert#1\right\rVert}
\newcommand{\mean}[1]{\left\langle#1\right\rangle_V}

\theoremstyle{plain}

\newtheorem{theorem}{Theorem}[section]
\begin{document}

\title[Self-consistent mixed thermal-runaway electron equilibria]{Self-consistent mixed thermal-runaway electron equilibria}

\author{M. Keu}
 \email{markusedison.keu@anu.edu.au}
\author{M. Hole}%
\affiliation{Mathematical Sciences Institute, The Australian National University, Canberra, ACT 0200, Australia}%

\author{A. Malaquias}
\affiliation{Instituto de Plasmas e Fusão Nuclear, Instituto Superior Técnico, Universidade de Lisboa, 1049-001 Lisboa, Portugal}%

\date{\today}

\begin{abstract}
In this work, we develop and solve a self-consistent equilibrium model for plasmas containing thermal and runaway electron (RE) populations, motivated by observations of residual plasma density during alternating-current (AC) transitions in ISTTOK. We adopt an existing monoenergetic, narrow-pitch guiding-center formulation, while retaining a spatially varying toroidal-field function, and show that prescribing the radius-weighted RE density by the orbit label alone conflicts with particle conservation and vector equilibrium wherever the thermal toroidal-field function varies along an occupied drift contour. Both conditions instead require the ratio of RE density to toroidal field to be constant along each drift contour. Taking this ratio as a function of the orbit label gives a compatible closure that relates the equilibrium RE density to a prescribed reference profile. Combined with Grad--Shafranov--Helmholtz thermal free functions, it yields a closed field equation that retains the displacement between drift and magnetic surfaces. At fixed thermal coefficients, we prove existence for every finite reference-profile amplitude under sign and spectral conditions. We also prove local constrained existence and uniqueness near a thermal equilibrium with a nonsingular current--beta response and derive a current--pressure bound. An ISTTOK ramp-down example with a prescribed 20\% energetic-current fraction and 2~keV energy gives a magnetic--orbit axis separation of 2.39~mm. The resulting fields and current distributions can serve as initial conditions for studies of electron survival and plasma recovery through AC reversal.
\end{abstract}
\maketitle
\raggedbottom
% ---- begin model.tex ----

\section{Introduction}\label{sec:intro}

Alternating-current (AC) tokamak operation was first demonstrated in STOR-1M, with residual plasma density through current reversal,\cite{Mitarai1987} and subsequently in JET at $\pm2$~MA with a finite dwell before renewed breakdown.\cite{Tubbing1992} CT-6B and HT-7 also achieved AC operation, while ISTTOK obtained multicycle discharges without dwell time.\cite{Huang2000,Li2007,Fernandes2002} In CT-6B, inversion of internal magnetic-probe measurements reconstructed oppositely directed currents at zero net plasma current. The loop voltage crossed zero about 0.5~ms before current reversal.\cite{Huang2000} In lower-hybrid-assisted HT-7 discharges, finite loop voltage was reported at zero plasma current. Grad--Shafranov--Helmholtz modeling yielded oppositely directed toroidal currents on the high- and low-field sides.\cite{Li2007}

In the ISTTOK experiments reported by Malaquias \textit{et al.}, residual plasma density persisted through coincident zero crossings of the net toroidal plasma current $I_p$ and the loop voltage, within the diagnostic resolution. Pressure-like profiles measured by the heavy ion beam diagnostic (HIBD) showed no resolved evidence of coexisting antiparallel current channels. These observations led to the hypothesis that energetic electrons remaining from the preceding half-cycle continue to ionize background neutrals, thereby helping to sustain the residual plasma.\cite{Malaquias2022} This interpretation also differs from the limiter short-circuit mechanism proposed for STOR-1M.\cite{Mitarai1992}

Runaway electrons (REs) arise when electric acceleration exceeds collisional drag for part of the electron distribution. Diffusion from the thermal tail can produce an initial RE population, which can then grow through avalanche multiplication driven by large-angle electron--electron collisions.\cite{Breizman2019} In toroidal geometry, magnetic curvature and field-gradient drifts displace their guiding-center orbits from magnetic-flux surfaces, with the displacement and confinement depending on electron momentum and the poloidal magnetic field.\cite{KnoepfelSpong1979} When REs carry an appreciable fraction of the plasma current, their contribution modifies the magnetic field, which in turn affects their drift orbits and the force balance of the thermal plasma. 

Knoepfel and Spong reviewed early self-consistent beam--plasma equilibria,\cite{KnoepfelSpong1979} while Yoshida calculated equilibria containing multiple electron beams with different energies, treating their drift orbits and magnetic fields together.\cite{Yoshida1990} The relativistic Hamiltonian theory developed by Tao, Chan, and Brizard provides a guiding-center description of charged-particle motion and its conserved quantities.\cite{Tao2007} More recently, Yuan and Hu developed a drift-surface solver for RE-current-dominated equilibria assuming a common parallel momentum,\cite{Yuan2023} while Bandaru and Hoelzl derived equilibrium formulations for plasmas containing both thermal and RE currents, explicitly accounting for relativistic drift-orbit effects and the displacement between drift surfaces and magnetic-flux surfaces.\cite{Bandaru2023}

We adopt Bandaru and Hoelzl's constant-energy, narrow-pitch reduction while retaining spatially varying $G=RB_\varphi$. Within these reduced equations, a density prescription $Rn_{\mathrm{RE}}=N(A)$ depending only on the orbit label $A=\gamma m_evR-e\psi$ conflicts with particle conservation and vector equilibrium where $F(\psi)$ varies along an occupied regular drift contour. The compatible relation $Rn_{\mathrm{RE}}=GQ(A)$ retains the particle-flux structure of their formulation A and relates a prescribed reference profile to the RE density in the mixed equilibrium.\cite{Bandaru2023}

\begin{figure*}[tbp]
\centering\includegraphics[width=5.5in]{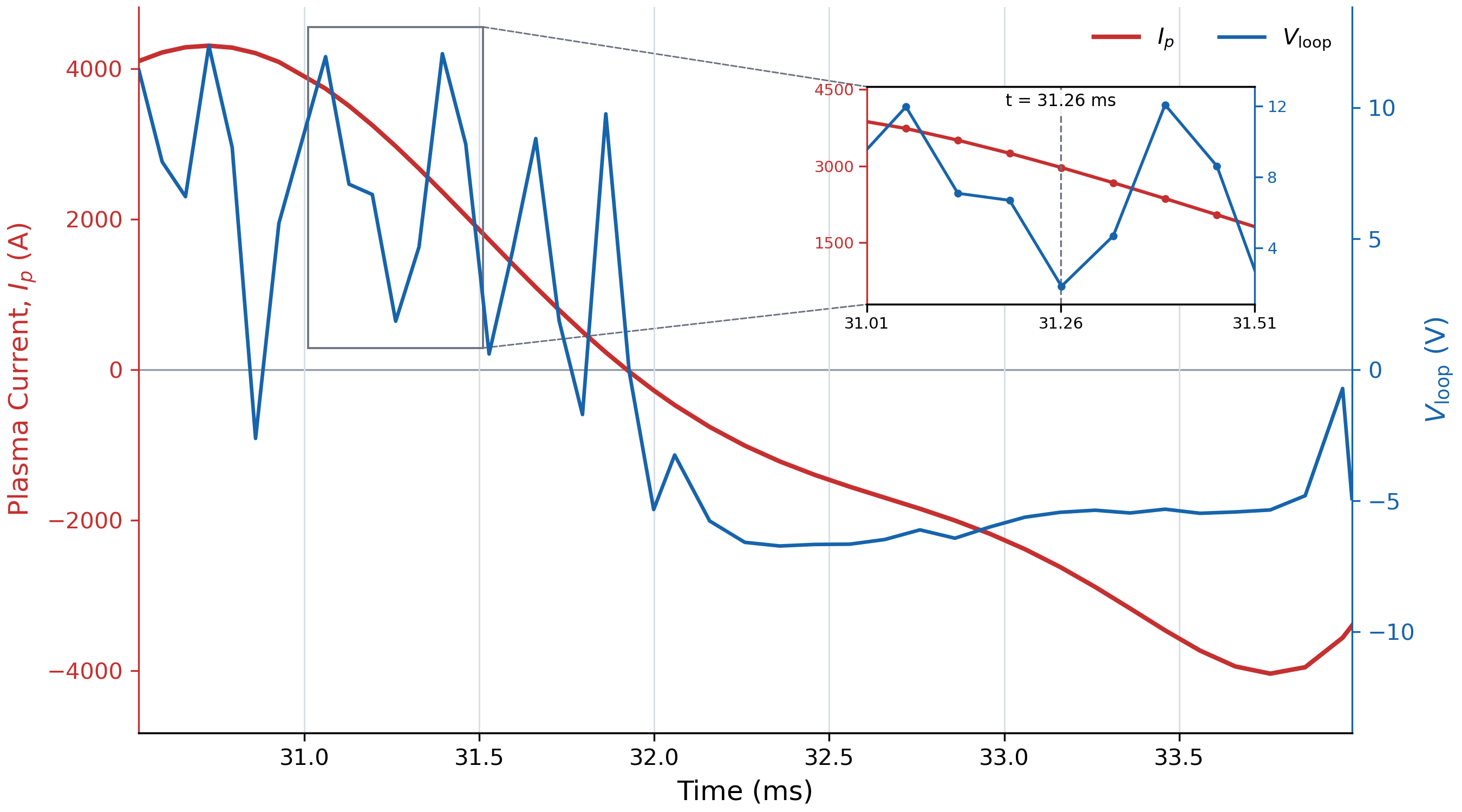}
\caption{Plasma current and loop voltage during the AC transition in ISTTOK discharge \#40041. The inset marks the equilibrium snapshot at 31.26~ms. The current crosses zero near 31.9~ms in this trace, about 0.65~ms after the snapshot.}\label{fig:context}
\end{figure*}

To describe the thermal component, we couple this density representation to the free functions introduced by Wang and further developed by Wang and Yu, obtaining a closed Grad--Shafranov--Helmholtz field equation.\cite{Wang2004,WangYu2005} The same thermal equilibrium family has also been applied to AC current reversal.\cite{Yu2006} Previous studies have established existence criteria and demonstrated the possibility of multiple solution branches for Grad--Shafranov problems.\cite{Jeyakumar2021,Ham2024} For the mixed thermal--RE family, we establish sufficient conditions for equilibrium and identify a current--pressure restriction on continuation toward reversal. We compare the pure-component limits with published solutions and examine spatial convergence. An ISTTOK ramp-down example (Fig.~\ref{fig:context}), with prescribed energetic-current fraction and energy, provides initial fields for studies of electron survival through current reversal.

Sections~\ref{sec:model}--\ref{sec:equilibrium_existence} present the model, closure and solvability results. Sections~\ref{sec:numerical_methods_results} and \ref{sec:ac_transition_discussion} give the numerical examples and their implications for AC reversal.

\section{Mixed thermal-RE equilibrium}\label{sec:model}

We consider an axisymmetric thermal plasma with scalar pressure and a directed RE population, following the constant-energy, narrow-pitch approximation of Bandaru and Hoelzl.\cite{Bandaru2023}

\subsection{Geometry, physical assumptions, and force balance}\label{sec:mixed_geometry}

We use cylindrical coordinates $(R,\varphi,Z)$, with $R$ the major radius, $\varphi$ the toroidal angle, and $Z$ the vertical coordinate. The plasma occupies the prescribed circular cross section
\begin{equation}
 \Omega=\{(R,Z):(R-R_0)^2+Z^2<a^2\},
 \label{eq:sec2_domain}
\end{equation}
where $R_0$ and $a$ are the major and minor radii, with $R_0>a>0$. In SI units, the axisymmetric magnetic field and Amp\`ere's law are
\begin{equation}
 \begin{aligned}
 \mathbf B&=\nabla\psi\times\nabla\varphi+G\nabla\varphi,\qquad G=RB_\varphi,\\
 \nabla\varphi&=\frac{\mathbf e_\varphi}{R}.
 \end{aligned}
 \label{eq:sec2_B}
\end{equation}
\begin{equation}
 \begin{aligned}
 \mu_0\mathbf J&=\nabla G\times\nabla\varphi-\Delta^*\psi\,\nabla\varphi,\\
 \Delta^*&=\partial_R^2-\frac{1}{R}\partial_R+\partial_Z^2.
 \end{aligned}
 \label{eq:sec2_ampere}
\end{equation}
Here $\psi$ is the poloidal magnetic flux divided by $2\pi$, $G$ is the toroidal-field function, and $\mathbf J$ is the total current density. We prescribe the circular boundary as a magnetic surface and choose its constant flux value to be $\psi_b=0$, with $G>0$. This fixes the additive flux origin without requiring a vanishing poloidal field: $\mathbf B_p=R^{-1}\nabla\psi\times\mathbf e_\varphi$ can remain finite because the normal derivative of $\psi$ need not vanish. The usual axis-zero, boundary-one normalization is $\psi_N=(\psi-\psi_a)/(\psi_b-\psi_a)=1-\psi/\psi_a$, where $\psi_a$ is the magnetic-axis flux. The dimensional field uses $\nabla\psi=(\psi_b-\psi_a)\nabla\psi_N$.

The thermal component has scalar pressure $p$, representing the combined pressure of the thermal species, and negligible bulk-flow inertia. In the stationary force balance of each species, we neglect electric forces and collisional momentum exchange. The REs have a single kinetic energy $E_k$ and a spatially constant signed momentum, with motion dominated by the toroidal component. With elementary charge $e>0$ and electron charge $-e$, we take
\begin{equation}
 \begin{aligned}
 \gamma&=1+\frac{E_k}{m_ec^2},\qquad v=-c\sqrt{1-\gamma^{-2}},\\
 K&=\gamma m_ev<0.
 \end{aligned}
 \label{eq:sec2_energy}
\end{equation}
Here $m_e$ is the electron rest mass and $c$ is the speed of light. The quantity $v$ is the signed parallel speed $v_{\parallel}$ in the toroidal-motion approximation. All REs share one kinetic energy and the same sign of $v_{\parallel}$, so
\begin{equation}
 \nabla\gamma=0,\qquad \nabla v_{\parallel}=\nabla v=0,
 \qquad \nabla K=0.
 \label{eq:sec2_constant_momentum}
\end{equation}
Uniform energy and a common sign of $v_{\parallel}$ are population assumptions, stronger than energy conservation along individual orbits. We assume $v_{\perp,\mathrm{gyro}}/|v_{\parallel}|\ll1$, retaining the guiding-center curvature drift. This pitch ordering is distinct from toroidal-field dominance, $B_p/B_\varphi\ll1$. The velocity vector varies with the geometry. The sign $v<0$ gives current in the positive prescribed plasma-current direction. We reserve $p$ for thermal pressure.

The electron orbit label and radius-weighted density are
\begin{equation}
 A=KR-e\psi,\qquad N=Rn_{\mathrm{RE}},
 \label{eq:sec2_A_N}
\end{equation}
where $n_{\mathrm{RE}}$ is the energetic-electron number density. Because $K$ is spatially constant,
\begin{equation}
 \nabla A=K\nabla R-e\nabla\psi.
 \label{eq:sec2_grad_A}
\end{equation}
The label $A$ has units of angular momentum, whereas $A/e$ has units of magnetic flux. An additive change $\widetilde\psi=\psi+C$ translates the label to $\widetilde A=A-eC$. Consistent relabelling of the free functions, reference label and population profile preserves the physical field and density.

In this narrow-pitch and monoenergetic approximation, the energetic-electron velocity is
\begin{equation}
 \begin{split}
 \mathbf v_{\mathrm{RE}}
 &=-\frac{v}{eG}\,\nabla A\times\mathbf e_\varphi+v\mathbf e_\varphi\\
 &=\frac{vR}{G}\,\mathbf B-\frac{vK}{eG}\,\mathbf e_Z.
 \end{split}
 \label{eq:sec2_vRE}
\end{equation}
The second form separates motion along the magnetic field from the vertical curvature drift. Since $\mathbf v_{\mathrm{RE}}\cdot\nabla A=0$, the guiding-center flow remains on surfaces of constant $A$ in this stationary model. On the magnetic boundary, however, $\mathbf v_{\mathrm{RE}}\cdot\widehat{\mathbf n}=-vK(\mathbf e_Z\cdot\widehat{\mathbf n})/(eG)$, where $\widehat{\mathbf n}$ is the outward normal. Curvature drift can therefore cross that surface. No reflecting or absorbing material-wall condition is imposed.

The curvature-drift current, $\mathbf J_{\mathrm{curv}}=n_{\mathrm{RE}}vK\mathbf e_Z/G$, provides the inward Lorentz force required for toroidal electron motion, as in Bandaru and Hoelzl~\cite{Bandaru2023}:
\begin{equation}
 \mathbf J_{\mathrm{curv}}\times(B_\varphi\mathbf e_\varphi)
 =-\frac{n_{\mathrm{RE}}vK}{R}\mathbf e_R
 =-\frac{n_{\mathrm{RE}}\gamma m_ev^2}{R}\mathbf e_R.
 \label{eq:sec2_curvature_balance}
\end{equation}
Subtracting the RE current leaves the thermal force balance,
\begin{equation}
 (\mathbf J-\mathbf J_{\mathrm{RE}})\times\mathbf B=\nabla p,
 \qquad \mathbf J_{\mathrm{RE}}=-e n_{\mathrm{RE}}\mathbf v_{\mathrm{RE}}.
 \label{eq:sec2_force_balance}
\end{equation}
Taking the scalar product with $\mathbf B$ shows that $p=p(\psi)$ on connected magnetic surfaces. The energetic momentum flux enters through the curvature force in Eq.~\eqref{eq:sec2_curvature_balance}. The pressure $p$ contains only the thermal contribution.

\subsection{Equilibrium and current-density equations}\label{sec:mixed_field_equations}

\begin{samepage}
Under these assumptions, the constant-energy form of formulation B in Ref.~\onlinecite{Bandaru2023} gives two equations for $\psi$ and $G$:
\begin{equation}
 \nabla G-F'(\psi)\nabla\psi
   =\mu_0v\frac{N}{G}\nabla A,
 \label{eq:sec2_vector_equilibrium}
\end{equation}
\begin{equation}
 \Delta^*\psi=-\mu_0R^2p'(\psi)-GF'(\psi)+\mu_0evN.
 \label{eq:sec2_scalar_equilibrium}
\end{equation}
Here $F(\psi)$ is the thermal toroidal-field free function, and a prime denotes differentiation with respect to $\psi$. The thermal function $F$ equals the full toroidal-field function $G=RB_\varphi$ in the absence of energetic electrons. 
\end{samepage}

The toroidal current separates into thermal and RE contributions,
\begin{equation}
 \begin{split}
 j_{\varphi,\mathrm{th}}&=Rp'(\psi)
                      +\frac{GF'(\psi)}{\mu_0R},\\
 j_{\varphi,\mathrm{RE}}&=-\frac{evN}{R},\qquad
 j_\varphi=j_{\varphi,\mathrm{th}}+j_{\varphi,\mathrm{RE}}.
 \end{split}
 \label{eq:sec2_species_currents}
\end{equation}

Since $v<0$ and $N\geq0$, the RE current is positive in our convention. The full field function $G$ enters the thermal current, so $GF'$ includes its response to the energetic population.

\subsection{Thermal free functions and global constraints}\label{sec:mixed_thermal}

To specify the thermal part of the equilibrium, we use the Grad--Shafranov--Helmholtz free functions introduced by Wang~\cite{Wang2004} and developed further by Wang and Yu~\cite{WangYu2005}. For a positive reference field $B_0$ and the boundary value $G_*=F(0)>0$ of the thermal field function, they are
\begin{equation}
 p'(\psi)=-\frac{B_0a_1}{\mu_0a^2},\qquad
 p(\psi)=p_b+p'(\psi)\psi,
 \label{eq:sec2_pressure_family}
\end{equation}
\begin{equation}
 F^2(\psi)=G_*^2+2B_0a_2\psi+\Lambda\psi^2,
 \qquad F(\psi)>0,
 \label{eq:sec2_F_family}
\end{equation}
\begin{equation}
 F(\psi)F'(\psi)=B_0a_2+\Lambda\psi,
 \qquad \Lambda=\frac{\alpha_{\mathrm{eff}}^2}{a^2}.
 \label{eq:sec2_F_derivative}
\end{equation}
The dimensionless coefficients $a_1$, $a_2$, and $\alpha_{\mathrm{eff}}$ control the thermal current and pressure profiles. The boundary thermal pressure is $p_b=p(0)$, and the positive field branch gives $F(0)=G_*$. Admissibility requires $F$ to remain real and positive throughout the plasma.

In the absence of REs, $N=0$ and $G=F$. Equation~\eqref{eq:sec2_scalar_equilibrium} then reduces to
\begin{equation}
 (\Delta^*+\Lambda)\psi
   =B_0\left[a_1\left(\frac{R}{a}\right)^2-a_2\right].
 \label{eq:sec2_thermal_GSH}
\end{equation}

This thermal Grad--Shafranov--Helmholtz equation recovers Wang's family. His circular Fourier--Bessel solution uses a large-aspect-ratio approximation. Wang and Yu's separated solution is exact for a rectangular cross section. Our circular-domain calculation retains the full toroidal operator $\Delta^*$, allowing a direct geometric comparison in Sec.~\ref{sec:pure_limits}.

We define the component currents by integrating over the poloidal cross section and the volume-averaged thermal beta using the reference field $B_0$:
\begin{equation}
 \begin{split}
 I_s&=\int_\Omega j_{\varphi,s}\,dS,
 \qquad s\in\{\mathrm{th},\mathrm{RE}\},\\
 \beta_V&=\frac{2\mu_0}{B_0^2}\langle p\rangle_V,
 \qquad
 \langle p\rangle_V
 =\frac{\int_\Omega Rp\,dS}{\int_\Omega R\,dS}.
 \end{split}
 \label{eq:sec2_global_constraints}
\end{equation}
Pressure is averaged with the toroidal-volume weight $R$, whereas each component current is integrated over poloidal area. By Amp\`ere's law, the total current equals the boundary poloidal-field circulation divided by $\mu_0$. The fixed toroidal field $B_0$ normalizes thermal beta, following Malaquias \emph{et al.}\cite{Malaquias2022}

For a positive total plasma current $I_p=I_{\mathrm{th}}+I_{\mathrm{RE}}$, we prescribe the energetic-current fraction $f_{\mathrm{RE}}=I_{\mathrm{RE}}/I_p$, together with $I_p$, $\beta_V$, $p_b$, and $E_k$. For a specified reference-profile shape and thermal profile parameter, $a_1$, $a_2$, and the RE amplitude are solved simultaneously with the field equation and global constraints, subject to admissibility. An independent constraint on central current density would require an additional free parameter.

For the ISTTOK example we use
\begin{equation}
 \alpha_{\mathrm{eff}}=\alpha_w\sqrt{1-f_{\mathrm{RE}}},
 \qquad 0\leq f_{\mathrm{RE}}<1,
 \label{eq:sec2_alpha_choice}
\end{equation}

where $\alpha_w$ is fixed before fitting. This modeling choice recovers $\alpha_{\mathrm{eff}}=\alpha_w$ in the thermal limit and makes $\Lambda=\alpha_w^2(1-f_{\mathrm{RE}})/a^2$ proportional to the thermal current fraction. Since $\Lambda\psi$ is only one term in $FF'$, the constraints still determine $a_1$ and $a_2$. Varying $f_{\mathrm{RE}}$ therefore changes both populations' contributions.

\section{RE population representation}\label{sec:re_representation}

\subsection{Compatible closure}\label{sec:re_compatible_closure}

The RE density must satisfy stationary particle conservation and vector equilibrium.

For a stationary RE population without local particle sources or losses, $\nabla\cdot(n_{\mathrm{RE}}\mathbf v_{\mathrm{RE}})=0$. Substituting Eq.~\eqref{eq:sec2_vRE}, with the spatially constant $v$ of Eq.~\eqref{eq:sec2_constant_momentum}, gives
\begin{equation}
 \nabla\cdot(n_{\mathrm{RE}}\mathbf v_{\mathrm{RE}})
 =-\frac{v}{e}\nabla\!\left(\frac{N}{G}\right)
 \cdot(\nabla A\times\nabla\varphi).
 \label{eq:sec3_continuity}
\end{equation}
Because $\nabla A\times\nabla\varphi$ is tangent to a contour of $A$, conservation requires $N/G$ to be constant along each connected, regular drift contour. The same condition follows by taking the curl of Eq.~\eqref{eq:sec2_vector_equilibrium}:
\begin{equation}
 \nabla\!\left(\frac{N}{G}\right)\times\nabla A=0.
 \label{eq:sec3_integrability}
\end{equation}
Both conditions are satisfied by choosing a single-valued $Q(A)$ and setting $N=GQ(A)$. Substitution into the vector equation gives
\begin{equation}
 \nabla[G-F(\psi)]=\mu_0vQ(A)\nabla A.
 \label{eq:sec3_integrated_vector}
\end{equation}
Integrating in $A$, with the additive constant absorbed into $F$, yields
\begin{equation}
 N=GQ(A),\qquad
 H_A=\mu_0vQ(A),\qquad
 G=F(\psi)+H(A).
 \label{eq:sec3_closure}
\end{equation}
Here $H(A)=G-F(\psi)$ is the orbit-dependent contribution to the toroidal-field function, and $H_A=dH/dA$. The ratio $Q=N/G=n_{\mathrm{RE}}/B_\varphi$ is constant along a drift contour, while the density varies as $n_{\mathrm{RE}}=GQ/R$. When $F$ varies along a drift contour, $N$ varies with $G$ so that $N/G$ remains constant. A single global $Q(A)$ assigns the same ratio $n_{\mathrm{RE}}/B_\varphi$ to disconnected contours with the same orbit label. Conservation requires constancy on each connected component, but not equality between different components. The common prescription is an additional closure assumption. It implies neither equal local densities nor particle transfer between the contours. At stationary points of $A$, these relations hold by continuous extension without division by $|\nabla A|$.

If $N=\mathcal N(A)>0$, Eq.~\eqref{eq:sec3_integrability} requires $\nabla G\times\nabla A=0$ on occupied regular contours. Crossing the vector equation with $\nabla A$ and using Eq.~\eqref{eq:sec2_grad_A} then gives
\begin{equation}
 K F'(\psi)\,\partial_Z\psi=0.
 \label{eq:sec3_obstruction}
\end{equation}
For $K\ne0$, this fails wherever $F'(\psi)\partial_Z\psi\ne0$ on an occupied contour. Pure-RE and mixed states with $F'=0$ avoid this obstruction (Appendix~\ref{app:density_obstruction}).

The closure retains the particle-flux stream-function structure of formulation A of Bandaru and Hoelzl.\cite{Bandaru2023} For constant energy, their stream function is $u(A)=-H(A)/(e\mu_0)$, giving $u_A=-vQ/e$ and $G=F-e\mu_0u$. Their uniform-energy prescription $N=N(A)$ follows at leading order when $G$ is approximately constant. The present closure enforces compatibility while retaining its spatial variation within the reduced equations.

\subsection{Reference profile and equilibrium density}\label{sec:re_reference}

We specify the energetic population through a reference profile $N_B(A)$. Its units, like those of $N$, are density multiplied by radius. The subscript $*$ denotes fixed reference values: $A_*$ is the label at which $H=0$, and $G_*>0$ is the reference field value. We define

\begin{equation}
 D_B(A)=G_*^2+2\mu_0v\int_{A_*}^{A}N_B(A')\,dA'.
 \label{eq:sec3_reference_rad}
\end{equation}

Where $D_B(A)>0$, we take the positive square root and set

\begin{equation}
 \begin{gathered}
 G_B(A)=\sqrt{D_B(A)},\qquad Q(A)=\frac{N_B(A)}{G_B(A)},\\
 H(A)=G_B(A)-G_*.
 \end{gathered}
 \label{eq:sec3_reference_functions}
\end{equation}

The quantity $D_B$ is the square of the reference toroidal-field function $G_B$. These definitions give $G_B(A_*)=G_*$ and $dH/dA=\mu_0vQ(A)$, satisfying Eq.~\eqref{eq:sec3_closure}. The corresponding radius-weighted density in the mixed equilibrium is

\begin{equation}
 N=\frac{G}{G_B}N_B
 =\left(1+\frac{F(\psi)-G_*}{G_B(A)}\right)N_B(A),
 \qquad n_{\mathrm{RE}}=\frac{N}{R}.
 \label{eq:sec3_actual_density}
\end{equation}

Thus $N=N_B$ when $F=G_*$. Otherwise, $G/G_B$ accounts for the thermal field contribution. Prescribing a Gaussian for $N_B$, $Q$, or $N$ gives different densities. Here we prescribe $N_B$ and obtain $Q$ and $N$ from Eqs.~\eqref{eq:sec3_reference_functions} and \eqref{eq:sec3_actual_density}.

Along a contour of fixed $A$, the relation $A=KR-e\psi$ gives $d\psi/dR=K/e$, and hence

\begin{equation}
 \left.\frac{dN}{dR}\right|_{A}
 =\frac{K}{e}\,Q(A)F'(\psi).
 \label{eq:sec3_along_contour}
\end{equation}

Here $R$ parametrizes a regular contour segment. The variation of $N$ along it is proportional to $F'(\psi)$. The number density also contains the geometric factor $1/R$.

\begin{samepage}
Substituting $G=F+H$ and $N=GQ$ into Eq.~\eqref{eq:sec2_scalar_equilibrium} with the Wang free functions of Sec.~\ref{sec:mixed_thermal} gives the closed field equation

\begin{equation}
 \begin{split}
 (\Delta^*+\Lambda)\psi={}&B_0\left[a_1\left(\frac{R}{a}\right)^2-a_2\right]\\
 &{}-H(A)F'(\psi)+\mu_0ev\,GQ(A).
 \end{split}
 \label{eq:sec3_mixed_GSH}
\end{equation}

Because $A=KR-e\psi$ and $G=F(\psi)+H(A)$ depend on the same $\psi$, the thermal and RE currents must be solved together.
\end{samepage}

\subsection{RE profile parameters and identifiability}\label{sec:re_parameter_identification}

We represent the reference profile $N_B$ by a Gaussian in the orbit-label coordinate $\zeta$:
\begin{equation}
 \zeta=\frac{\varepsilon_A(A-A_a)}{e},\qquad
 N_B(A)=N_0\exp\!\left[-\frac{(\zeta-d)^2}{2s^2}\right].
 \label{eq:sec3_gaussian}
\end{equation}
Here $A_a$ is the value of $A$ at a selected orbit axis, an interior extremum of $A$. We choose $\varepsilon_A=\pm1$ so that $\zeta$ increases outward from that axis. The parameter $N_0\geq0$ is the amplitude of the radius-weighted reference profile, $d\geq0$ locates its peak relative to the orbit axis, and $s>0$ is its width in $\zeta$. Thus $d$ and $s$ have units of magnetic flux (Wb), as do $\zeta$ and $\psi$.

\begin{samepage}
We compare reference shapes using two equivalent descriptors. With $g(\zeta)=\exp[-(\zeta-d)^2/(2s^2)]$, we define
\begin{equation}
 \begin{split}
 X_h&=\exp\!\left(-\frac{d^2}{2s^2}\right),\\
 M_{2,A}&=\frac{\displaystyle\int_0^{\infty}\zeta^2g(\zeta)\,d\zeta}
 {\displaystyle\int_0^{\infty}g(\zeta)\,d\zeta}.
 \end{split}
 \label{eq:sec3_shape_parameters}
\end{equation}
\end{samepage}
For $N_0>0$, $X_h=N_B(A_a)/N_0$ is the reference axis-to-peak ratio: it is one for an axis-centered profile and decreases with $d/s$. The second moment $M_{2,A}$, in Wb$^2$, measures breadth about the orbit axis over all $\zeta\geq0$, including labels outside the plasma's attained range. Its square root is the root-mean-square value of $\zeta$ under the reference profile.

With $\phi_N$ and $\Phi$ denoting the standard normal probability density and cumulative distribution functions, the inverse map is

\begin{equation}
 \begin{gathered}
 k=\sqrt{-2\ln X_h},\qquad
 D(k)=1+k^2+k\frac{\phi_N(k)}{\Phi(k)},\\[2pt]
 s=\sqrt{\frac{M_{2,A}}{D(k)}},\qquad d=ks.
 \end{gathered}
 \label{eq:sec3_shape_inverse}
\end{equation}

For $0<X_h\leq1$ and $M_{2,A}>0$, Eq.~\eqref{eq:sec3_shape_inverse} gives a one-to-one relation between the two parameter pairs (Appendix~\ref{app:descriptors}). Direct parameterization by $(d,s)$ avoids the nondifferentiability of the inverse map at $X_h=1$, where $d=0$.

If $A_a$ follows the equilibrium orbit axis, it changes with the magnetic field. Fixing $A_a$ defines a different reference-profile choice.

Although $(X_h,M_{2,A})$ uniquely determines $(d,s)$, the coupled field and thermal response can give similar total-current profiles for different reference shapes. Inferring the shape requires sufficiently sensitive observations. The energetic-current fraction $f_{\mathrm{RE}}$ and energy $E_k$ remain prescribed. At $N_0=0$, the field is independent of the Gaussian shape.

% ---- end model.tex ----
% ---- begin analysis_method.tex ----
\section{Solvability and current-pressure bound}
\label{sec:equilibrium_existence}

The closure enforces particle conservation and vector integrability. Solvability of the nonlinear scalar equation is established below, first at fixed thermal coefficients and then locally under prescribed integral constraints. Appendix~\ref{app:existence} gives the proofs.

\subsection{Existence at fixed thermal coefficients}
\label{sec:field_existence}

We fix the thermal coefficients $a_1$, $a_2$, $\Lambda$, the electron momentum, and a nonnegative reference shape $\chi(A)$, including its dimensional center and width. We write $N_B(A)=\eta N_{\mathrm{ref}}\chi(A)$, where $\eta\geq0$, $\chi$ is dimensionless, and $N_{\mathrm{ref}}>0$ is a fixed radius-weighted density scale (m$^{-2}$). The reference values remain fixed as $\eta$ varies. We denote by $\lambda_1(P)$ the lowest eigenvalue of $P=-\Delta^*$ with homogeneous Dirichlet boundary conditions in $L^2(\Omega,dS/R)$. The existence result holds under
\begin{equation}
 a_1\leq0\leq a_2,\qquad (a_1,a_2)\ne(0,0),\qquad
 0\leq\Lambda<\lambda_1(P),
 \label{eq:sec4_existence_conditions}
\end{equation}
and the regularity, sign and decay conditions of Appendix~\ref{app:finite_existence}. A Gaussian satisfies the profile conditions for $A_*\geq K(R_0-a)$.

For every finite $\eta$, the nonlinear source $\mathcal T_\eta=HF'+\mu_0e|v|N$ is nonnegative and uniformly bounded over nonnegative fluxes. The positive inverse of $P-\Lambda$ supplies lower and upper solution bounds. A fixed-point argument between these bounds gives a classical solution of Eq.~\eqref{eq:sec3_mixed_GSH}, with $\psi>0$ inside and positive $F$, $G_B$ and $G$ (Appendix~\ref{app:finite_existence}). For $\eta>0$, the RE current is nonzero if the reference profile is positive on an interval of accessible orbit labels. Here $I_p$, $\beta_V$ and $f_{\mathrm{RE}}$ are outputs. This result does not imply uniqueness or guarantee a solution for prescribed values of these quantities. Physical validity still requires the reduced-model ordering.

\subsection{Local existence and uniqueness at prescribed current and thermal beta}
\label{sec:constrained_uniqueness}

At prescribed $I_p$, $\beta_V$ and $f_{\mathrm{RE}}$, the unknowns are $\psi$, $a_1$, $a_2$ and the reference amplitude. A nonsingular current--beta response allows the thermal coefficients to maintain both constraints as RE current is introduced. At a thermal equilibrium, the implicit-function argument requires an invertible field linearization and a nonsingular $2\times2$ current--beta response matrix, including the field response to $a_1,a_2$. The latter condition is independent of the field spectral gap (Appendix~\ref{app:thermal_feasibility}). With smooth prescriptions, positive field functions and a reference profile occupying nonzero area, a unique nearby branch exists for sufficiently small positive $f_{\mathrm{RE}}$ at fixed $I_p$ and $\beta_V$. A center following an orbit extremum is allowed if its selection depends smoothly on the field (Appendix~\ref{app:local}).

This uniqueness is local for the specified profile and energy. The theorem does not give a quantitative interval in $f_{\mathrm{RE}}$ or establish continuation to the $f_{\mathrm{RE}}=0.20$ example, whose reference center follows the orbit axis.

\subsection{A current--pressure bound}
\label{sec:pressure_restriction}

For the constant pressure derivative $c_p=p'(\psi)\geq0$, $F'\geq0$, $G>0$, and nonnegative thermal and energetic toroidal current densities, Appendix~\ref{app:pressure} gives
\begin{equation}
 \beta_V-\beta_b\leq C_{\mathrm{mix}}(1-f_{\mathrm{RE}})I_p^2,
 \qquad \beta_b=\frac{2\mu_0p_b}{B_0^2},
 \label{eq:sec4_current_pressure_bound}
\end{equation}
where $C_{\mathrm{mix}}>0$ depends on geometry and $B_0$. The thermal current limits the pressure gradient, while the total current bounds the flux. Their product bounds excess thermal beta. Thus $\beta_V-\beta_b\to0$ as $I_p\to0$ in this sign regime, although finite boundary pressure may remain.

For $R_0=0.46$~m, $a=0.085$~m and $B_0=0.47$~T, $C_{\mathrm{mix}}\leq7.63\times10^{-11}$~A$^{-2}$. At $I_p\simeq2971$~A and $f_{\mathrm{RE}}=0.20$, the ceiling is $5.39\times10^{-4}$, above the prescribed $\beta_V\simeq7.86\times10^{-5}$ at $p_b=0$.

\section{Numerical methods and examples}
\label{sec:numerical_methods_results}

We test the pure-component limits before fitting the mixed ISTTOK reference profile at prescribed current, thermal beta, energetic-current fraction and energy.

\subsection{Numerical solution of the coupled equilibrium}
\label{sec:num_iteration}

On the circular cross section, we use polar coordinates $(r,\theta)$:
\begin{equation*}
R=R_0+ar\cos\theta,\qquad Z=ar\sin\theta,\qquad0\leq r\leq1.
\end{equation*}
The full toroidal operator is
\begin{equation}
\begin{split}
a^2\Delta^*={}&\partial_r^2+
\left(\frac1r-\frac{a\cos\theta}{R}\right)\partial_r
+\frac1{r^2}\partial_\theta^2\\
&+\frac{a\sin\theta}{Rr}\partial_\theta.
\end{split}
\label{eq:sec5_operator}
\end{equation}
Centered finite differences use radial cell centers and periodic angular points. Regularity is imposed at the origin, and an outer ghost-cell relation enforces $\psi=0$. Current integrals use $dS=a^2r\,dr\,d\theta$, and volume averages include the additional factor $R$.

Wang's modal thermal solution provides the initial flux, after which the mixed iteration uses the full operator in Eq.~\eqref{eq:sec5_operator}. For each flux iterate, we evaluate $A$, locate the orbit-axis label $A_a$, and update $N_B$, $G_B$, $H$, $F$, $G=F+H$ and $N=(G/G_B)N_B$. The thermal coefficients and amplitude satisfy
\begin{equation*}
I_{\mathrm{th}}=(1-f_{\mathrm{RE}})I_p,\qquad
I_{\mathrm{RE}}=f_{\mathrm{RE}}I_p,\qquad\beta_V=\beta_V^{\mathrm{input}}.
\end{equation*}
Here $g(A)$ denotes the unit-amplitude Gaussian in Eq.~\eqref{eq:sec3_gaussian} evaluated at $\zeta(A)$. At fixed $\psi$, $F$, and profile shape, writing $N_B=N_0g(A)$ gives the amplitude equation
\begin{equation}
-ev\int_\Omega\frac{N_0g(A)}R
\left[1+\frac{F(\psi)-G_*}{G_B(A;N_0)}\right]dS=f_{\mathrm{RE}}I_p.
\label{eq:sec5_current_norm}
\end{equation}
Appendix~\ref{app:amplitude} proves that this normalization has a unique nonnegative amplitude under the sign and profile assumptions stated there. Since $G_B$ depends on $N_0$, the equation is solved again whenever the field or reference center changes.

The linear field update is relaxed according to
\begin{equation*}
\psi^{(k+1)}=(1-\omega_k)\psi^{(k)}+\omega_k\widehat\psi^{(k+1)}.
\end{equation*}
The nominal calculation uses Aitken acceleration from $\omega_0=0.55$ and converges in 13 iterations. Appendix~\ref{app:refinement_details} lists the stopping criteria. Section~\ref{sec:numerical_accuracy} distinguishes the scalar residual from numerical differentiation defects in the analytical closure identities.

Admissible states have positive $F$, $G_B$ and $G$, nonnegative density and pressure, and the assumed current signs. Fits exclude trials that fail these conditions or the convergence criteria.

In this work, ChatGPT 5.6 Sol (OpenAI) and Claude Opus 5.5 (Anthropic) were used to independently check the derivations and numerical results for errors. We defined the scientific road map, carried out all derivations and proofs, developed the numerical code, performed the calculations, and verified every issue the tools raised.

\subsection{Pure-component limits and reference comparisons}
\label{sec:pure_limits}
% ---- begin reference_benchmarks.tex ----
\subsubsection{Pure-thermal limit}
Setting $N_B=0$ gives $H=0$, $G=F$ and Eq.~\eqref{eq:sec2_thermal_GSH}. We test this limit against Wang's large-aspect-ratio Fourier--Bessel expansion for the normal configuration in his Fig.~1.\cite{Wang2004} Both calculations use $R_0/a=5$, $a_1=-0.04531$, $a_2=-1.0808$ and $\alpha_w=2.337304$, with $\psi=0$ at the circular computational boundary. We express this dimensionless example in SI units using $a=1$~m, $R_0=5$~m and $B_0=1$~T.

With $x=R/a$, $y=Z/a$, $x_0=R_0/a$, $u=\psi/(B_0a^2)$ and $\upsilon=u/x$, the full thermal equation becomes
\begin{equation}
\left(\partial_x^2+\partial_y^2+\frac1x\partial_x-\frac1{x^2}
+\alpha_w^2\right)\upsilon=a_1x-\frac{a_2}{x}.
\label{eq:sec5_wang_full}
\end{equation}
Wang's large-aspect-ratio reduction replaces $x^{-1}\partial_x$ by $x_0^{-1}\partial_x$ and omits $-\upsilon/x^2$. We evaluate his Fourier--Bessel expansion with 28 radial modes and angular harmonics $0\leq m\leq10$. We solve Eq.~\eqref{eq:sec2_thermal_GSH} on a $256\times256$ grid with the full toroidal operator used for the mixed model. At $N_B=0$ and fixed coefficients, this is a linear Dirichlet problem.

For each computed flux, current and pressure are evaluated from the same thermal free functions,
\begin{equation*}
\begin{split}
j_\phi&=\frac{B_0}{\mu_0a}\left[-a_1x+\frac{a_2+\alpha_w^2u}{x}\right],\\
p&=-\frac{B_0^2a_1}{\mu_0}u,\qquad p_b=0.
\end{split}
\end{equation*}
The current plotted for Wang's solution is evaluated from these free functions. Applying the full toroidal Amp\`ere operator to his approximate flux gives a different current because that flux has a nonzero residual in the full equation. Thermal beta uses the volume weight $R\,dS$ in Eq.~\eqref{eq:sec2_global_constraints}.

\begin{figure*}[!t]
\centering
\includegraphics[width=0.64\textwidth]{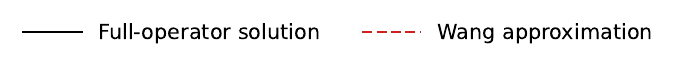}\\[3pt]
\begin{minipage}[b]{0.32\textwidth}\centering
\includegraphics[width=\linewidth]{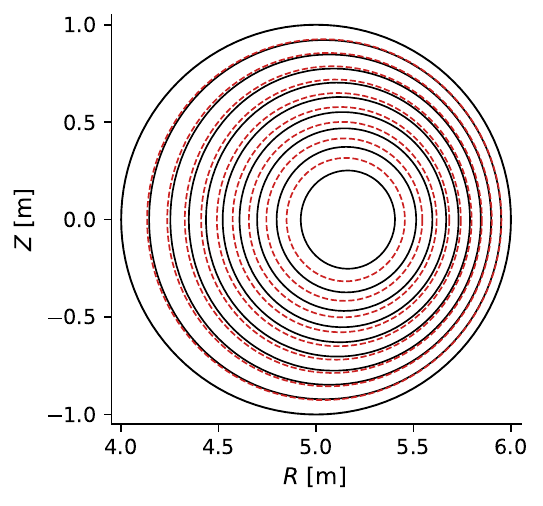}\\[2pt]
\small (a) Magnetic-flux surfaces
\end{minipage}\hfill
\begin{minipage}[b]{0.32\textwidth}\centering
\includegraphics[width=\linewidth]{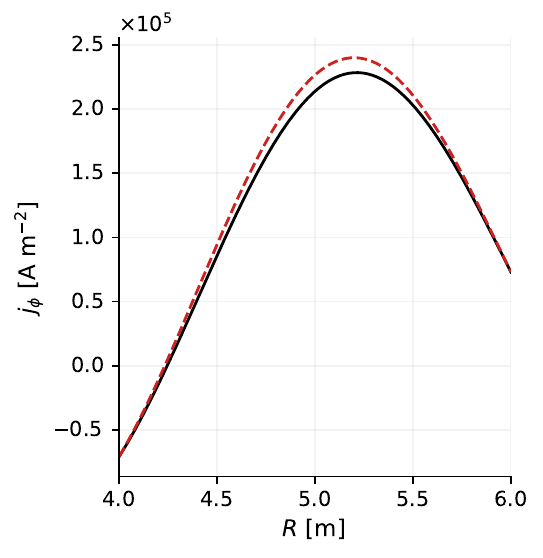}\\[2pt]
\small (b) Current density
\end{minipage}\hfill
\begin{minipage}[b]{0.32\textwidth}\centering
\includegraphics[width=\linewidth]{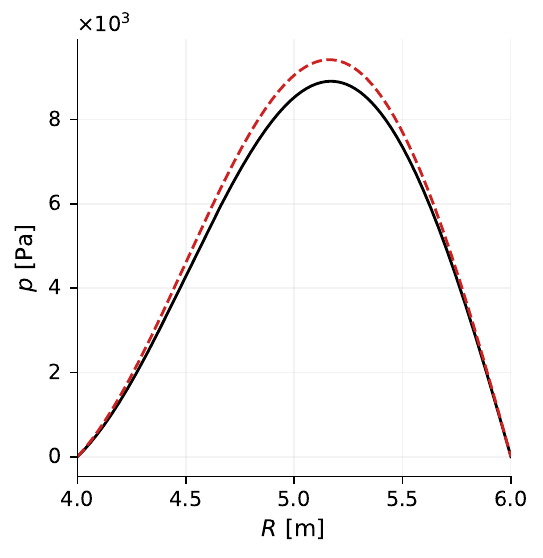}\\[2pt]
\small (c) Thermal pressure
\end{minipage}
\caption{Pure-thermal comparison at Wang's published coefficients: the full-operator solution (black solid) and Wang's large-aspect-ratio Fourier--Bessel expansion (red dashed). The dimensional parameters are $a=1$~m, $R_0=5$~m and $B_0=1$~T. Panel (a) overlays the same physical flux levels $\psi=0.025,0.050,\ldots,0.225$~Wb, with the circular computational boundary in black. Panels (b) and (c) show current density and thermal pressure at $Z=0$, evaluated from the common free functions.}
\label{fig:wang_benchmark}
\end{figure*}

% ---- begin thermal_comparison_results.tex ----
For a field $q$, the relative difference over the poloidal cross section is
\begin{equation*}
D_q=\frac{\|q_{\mathrm{full}}-q_W\|_{2,S}}{\|q_W\|_{2,S}},
\qquad \|q\|_{2,S}^2=\int_\Omega |q|^2\,dS,
\end{equation*}
where ``full'' denotes the full toroidal operator and $W$ denotes Wang's approximation. We obtain $D_\psi=D_p=5.57\%$ and $D_{j_\phi}=5.06\%$. Pressure and flux have the same relative difference because their proportionality coefficient is fixed. The midplane relative current difference is $5.17\%$, and the volume-averaged thermal beta is $5.50\%$ lower with the full operator (Table~\ref{tab:thermal_comparison} and Fig.~\ref{fig:wang_benchmark}).

\begin{table*}[tbp]
\centering\small
\caption{Fixed-coefficient pure-thermal comparison in Fig.~\ref{fig:wang_benchmark}. The final column gives the signed difference (full operator minus Wang), in the units of the first column. All integrals use the same grid. Values are rounded.}
\label{tab:thermal_comparison}
\begin{tabular*}{\textwidth}{@{\extracolsep{\fill}}lrrr@{}}\toprule
Quantity & Full-operator solution & Wang approximation & Difference\\\midrule
Volume-averaged thermal beta $\beta_V$ & 0.009569 & 0.010125 & $-0.000557$\\
Total current $I_p$ (kA) & 301.02 & 317.81 & $-16.80$\\
Magnetic-axis flux $\psi_a$ (Wb) & 0.24705 & 0.26124 & $-0.01420$\\
Magnetic-axis radius $R_a$ (m) & 5.16932 & 5.16132 & $+0.00800$\\
Peak midplane current density (kA~m$^{-2}$) & 228.38 & 239.99 & $-11.61$\\
Peak midplane thermal pressure (kPa) & 8.908 & 9.419 & $-0.512$\\\bottomrule
\end{tabular*}
\end{table*}

At the nine common physical flux levels, the maximum contour distance ranges from 10.3 to 73.0~mm, reaching $7.30\%$ of the minor radius. These distances include changes in flux amplitude and axis position. The approximately $5.6\%$ flux difference persists under refinement of both methods and reflects the geometric approximations and modal projection in Wang's expansion. Appendix~\ref{app:thermal_comparison_details} defines the contour metric and gives the numerical checks.
% ---- end thermal_comparison_results.tex ----

\subsubsection{Pure-RE limit}
Setting the thermal pressure and current contributions to zero pointwise, $p'=F'=0$, $F=G_*$ and $p_b=0$, gives $G=G_B$ and $N=N_B$. Applying $\Delta^*$ to $A=KR-e\psi$ yields
\begin{equation}
\Delta^*A=-\frac KR-\mu_0e^2vN_B(A),\qquad
A\big|_{\partial\Omega}=KR.
\label{eq:sec5_pure_re}
\end{equation}
We compare with the circular example of Bandaru and Hoelzl,\cite{Bandaru2023} using $R_0=6.2$~m, $a=2$~m, $B_0=5.3$~T, $I_p=I_{\mathrm{RE}}=10^7$~A and energies of 40 and 80~MeV. We set $\psi=0$ on the computational boundary and take $A_*/e=0$ and $G_*=32.86$~T~m.

For this calculation, the reference Gaussian is
\begin{equation}
N_B(A)=N_0\exp\!\left[-\frac{(A/e-\zeta_c)^2}{0.05\ell_A^2}\right],
\qquad \zeta_c=A_a/e,
\label{eq:sec5_pure_gaussian}
\end{equation}
where $A_a$ is the label at the chosen orbit extremum, recalculated from each flux iterate, and $\ell_A=\max_{\partial\Omega}|A/e-\zeta_c|$. The Gaussian peaks at $A/e=\zeta_c$, and the dimensionless width $0.05$ corresponds to the dimensional variance $s^2=0.025\ell_A^2$. The current constraint determines $N_0$ at each iterate. Table~\ref{tab:re_benchmark} gives the resulting profile parameters.

The labels ``Constant Radius'' and ``Local Radius'' refer to the radius factor in the RE current contribution to the scalar field equation. For Constant Radius, $N_B=R_{\mathrm{ref}}n_0g(A)=N_0g(A)$, where the fixed factor $R_{\mathrm{ref}}$ is absorbed into $N_0$. For Local Radius, $N=Rn_0g(A)$, so this prefactor varies along an orbit contour. Both calculations retain the local $R$ in the geometry, orbit label and curvature term. The Local Radius calculation solves the scalar field equation. It violates the pure-RE vector compatibility condition where its radius prefactor varies along an occupied drift contour, as shown in Appendix~\ref{app:source_weightings}.

Figure~\ref{fig:bandaru_benchmark} compares Constant Radius (black solid) and Local Radius (red dashed) on $256\times256$ grids. An independent scalar calculation of the Constant Radius prescription agrees with the coupled solver in the area-weighted flux norm to $2.95\times10^{-6}$ at 40~MeV and $3.21\times10^{-6}$ at 80~MeV. The independent scalar calculation reconstructs $G_B$ using $G_B(K(R_0-a))=G_*$, whereas the coupled calculation uses $G_B(0)=G_*$. At fixed $N_B$, this reference choice changes only the reconstructed toroidal field and does not enter Eq.~\eqref{eq:sec5_pure_re}.

The boundary-contact orbit contour has label $A_{\mathrm{bc}}=\min_{\partial\Omega}A$ and is tangent to the outboard computational boundary on this interior-minimum branch. The boundary radii are $R_-=R_0-a$ and $R_+=R_0+a$. The radius $R_{\mathrm{in}}$ is the opposite midplane crossing of $A(R,0)=A_{\mathrm{bc}}$. The contour's center displacement is
\begin{equation}
\delta=\frac{R_{\mathrm{in}}+R_+}{2}-R_0.
\label{eq:sec5_shift_metric}
\end{equation}
The opposite crossing is found by bracketed root finding on a cubic spline through the midplane values and the prescribed boundary values. The displacement $\delta$ locates the center of the boundary-contact orbit contour. The Constant Radius shifts are $0.07368$ and $0.13833$~m, while the Local Radius shifts are $0.08211$ and $0.15258$~m. The latter are closer to the approximate published values of $0.08$ and $0.15$~m. The computed maximum $B_p/B_\phi$ is about $0.81$ in these high-energy examples. Section~\ref{sec:ac_transition_discussion} discusses the implications for the toroidal-motion approximation.

\begin{table*}[tbp]
\centering\small
\caption{Pure-RE Gaussian specifications and boundary-contact contour shifts on $256\times256$ grids. For Constant Radius, the amplitude is $N_0$ in m$^{-2}$. For Local Radius it is $n_0$ in m$^{-3}$. Values are rounded for display.}
\label{tab:re_benchmark}
\begin{tabular}{@{}clrrrrr@{}}\toprule
$E_k$ (MeV)&Source prescription&$\zeta_c$ (Wb)&$\ell_A$ (Wb)&$2s^2$ (Wb$^2$)&Amplitude&$\delta$ (m)\\\midrule
40&Constant Radius&$-28.4252$&27.8577&38.8025&$3.60446\times10^{18}$&0.07368\\
80&Constant Radius&$-29.1032$&27.9753&39.1308&$3.48705\times10^{18}$&0.13833\\
40&Local Radius&$-28.9398$&28.3723&40.2495&$5.77486\times10^{17}$&0.08211\\
80&Local Radius&$-29.6040$&28.4761&40.5445&$5.56677\times10^{17}$&0.15258\\\bottomrule
\end{tabular}
\end{table*}

\begin{figure*}[tbp]
\centering\includegraphics[width=0.86\textwidth]{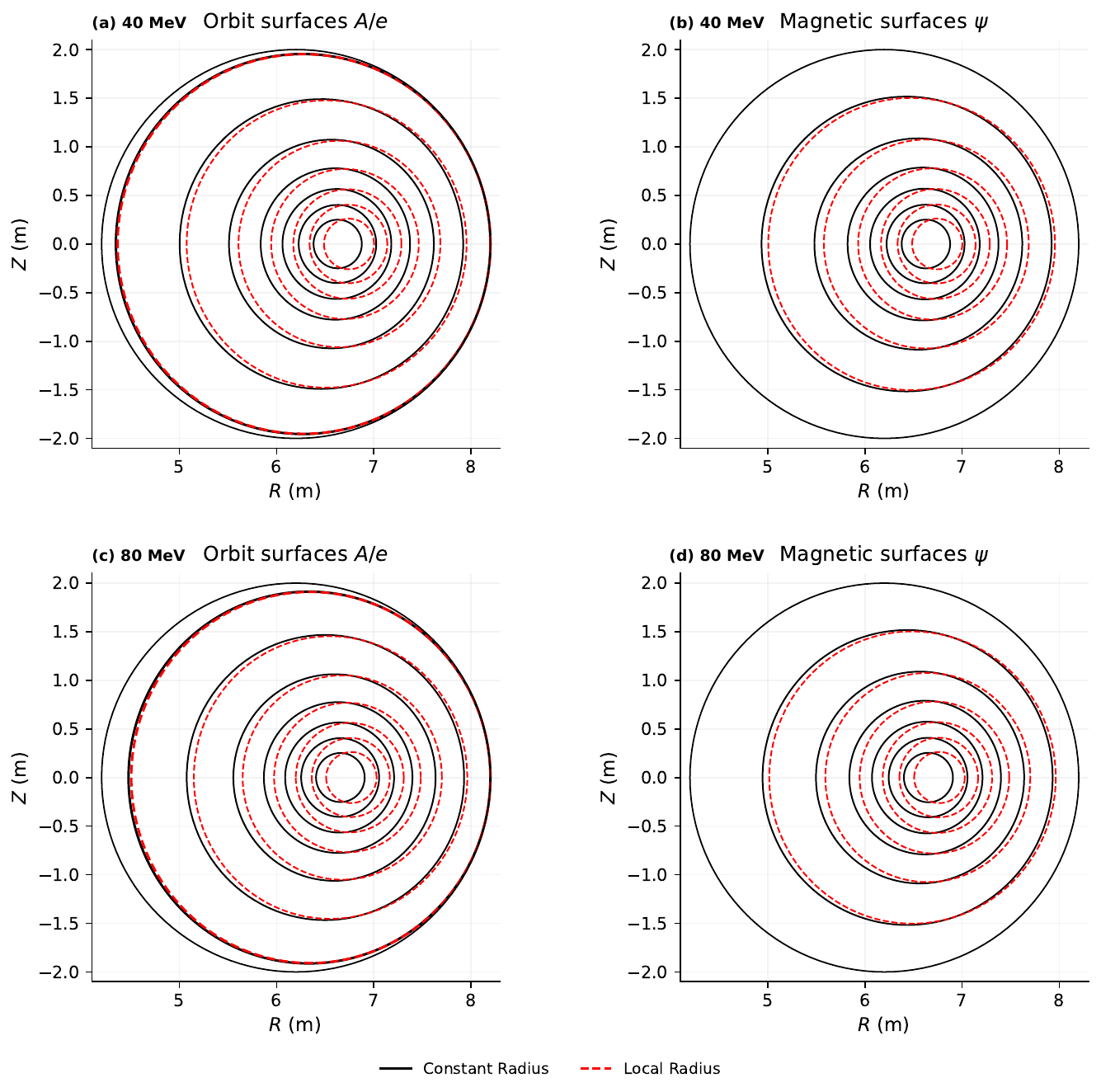}
\caption{Pure-RE scalar-equation comparisons for Constant Radius (black solid) and Local Radius (red dashed) at 40~MeV (top row) and 80~MeV (bottom row). Panels (a) and (c) show orbit-label contours $A/e$, including the boundary-contact orbit contour. Panels (b) and (d) show magnetic-flux contours $\psi$. Each panel uses common physical contour levels for the two prescriptions. The black circle is the prescribed computational boundary. Constant Radius and Local Radius differ only in the radius prefactor of the RE density, as defined in Sec.~\ref{sec:pure_limits}.}
\label{fig:bandaru_benchmark}
\end{figure*}
% ---- end reference_benchmarks.tex ----

\subsection{Fitting the reference-profile shape}
\label{sec:conditional_fitting}

Each trial $(X_h,M_{2,A})$ specifies $(d,s)$ through Eq.~\eqref{eq:sec3_shape_inverse}. We re-solve the constrained equilibrium, including the thermal coefficients, amplitude and orbit-axis reference.

For the estimated current-profile samples $j^{\mathrm{rec}}_{\phi,i}$, we define
\begin{equation}
\widehat j_i=\mathcal H_i[j_\phi],\qquad
E_j=\frac{\left[\sum_i(\widehat j_i-j^{\mathrm{rec}}_{\phi,i})^2\right]^{1/2}}
{\left[\sum_i(j^{\mathrm{rec}}_{\phi,i})^2\right]^{1/2}},
\label{eq:sec5_Ej}
\end{equation}
and
\begin{equation}
S(X_h,M_{2,A})=\sum_i\left(\frac{\widehat j_i-j^{\mathrm{rec}}_{\phi,i}}{s_j}\right)^2,
\qquad s_j=\max_i|j^{\mathrm{rec}}_{\phi,i}|.
\label{eq:sec5_S}
\end{equation}
The scale $s_j$ is the largest sample, so $S$ and $E_j$ have the same minimizers. The operator $\mathcal H_i$ interpolates the model total current linearly at $Z=0$ to $R_i=0.375+0.010i$~m, $i=0,\ldots,17$. The vertical diagnostic profile is compared with this midplane cut by assuming $\Delta R=R-R_0$ for the signed sampling coordinate.

The HIBD-derived $n_e\sigma_{\mathrm{ion}}$ profile, with ionization cross section $\sigma_{\mathrm{ion}}$,\cite{Malaquias2022,Malaquias2015} is represented by a polynomial preserving its average. Division by an assumed density distribution scaled using interferometry gives $\sigma_{\mathrm{ion}}$, from which $T_e$ and the Spitzer resistivity $\eta_{\mathrm{Sp}}$ are inferred. The preliminary current $\widetilde j_i=E_0/\eta_{\mathrm{Sp}}(T_{e,i})$ is normalized as $j^{\mathrm{rec}}_{\phi,i}=I_p\widetilde j_i/\mathcal Q_{\mathrm{rec}}[\widetilde j]$, using a half-annular area quadrature $\mathcal Q_{\mathrm{rec}}$ and one factor preserving both sides' relative amplitudes.

Twelve positions are HIBD-supported. Four linear extensions and two zero endpoints enter the initial $n_e\sigma_{\mathrm{ion}}$ profile. Their final currents follow the temperature reconstruction. Prescribed endpoint temperatures of $10^{-3}$~eV give negligible currents of $0.145$~A~m$^{-2}$, assuming connection to a zero-current exterior. All eighteen samples receive equal weight.

% ---- begin optimization_record.tex ----
The search covers $10^{-4}\leq X_h\leq1$ and $2\times10^{-10}\leq M_{2,A}\leq8\times10^{-9}$~Wb$^2$. A Sobol design followed by bounded least-squares searches on a $64^2$ grid gives a fitted pair near $X_h=0.5233$, $M_{2,A}=2.8175\times10^{-9}$~Wb$^2$, with $S=0.31257$. The six local searches give objective values agreeing within $1.6\times10^{-9}$. Appendix~\ref{app:search_details} gives the search parameters, sensitivity analysis, and profiled misfit curves.

Local refitting gives $X_h=0.5239$ and $M_{2,A}=2.8188\times10^{-9}$~Wb$^2$ on the $128^2$ grid, and $X_h=0.5241$ and $M_{2,A}=2.8194\times10^{-9}$~Wb$^2$ on the $256^2$ grid. The resulting $256^2$ solution is the nominal equilibrium. Its fitted descriptor pair is held fixed in the field-refinement study.
% ---- end optimization_record.tex ----

\subsection{ISTTOK example}
\label{sec:isttok_example}

We consider ISTTOK discharge \#40041 at $t=31.26$~ms during positive-current ramp-down, about $0.65$~ms before the current zero in Fig.~\ref{fig:context}. Table~\ref{tab:isttok_inputs} lists the inputs. The prescribed $f_{\mathrm{RE}}=0.20$ and $E_k=2$~keV illustrate a mixed equilibrium at this timestamp. The energy lies within the few-hundred-eV to few-keV range considered in the AC-transition analysis of Malaquias \emph{et al.}\cite{Malaquias2022} It gives $\gamma\simeq1.0039$ and $|v|\simeq0.0882c$. Neither input is inferred from the diagnostics, and the assigned energy alone does not establish runaway status. Generation and survival require a time-dependent calculation.

\begin{table*}[tbp]
\begin{minipage}[t]{0.47\textwidth}
\centering\small
\caption{Prescribed ISTTOK inputs and derived current shares. The total thermal pressure is approximated by the estimated electron pressure, with ion pressure neglected.}
\label{tab:isttok_inputs}
\begin{tabular}{@{}ll@{}}\toprule
Quantity & Value\\\midrule
$R_0$, $a$ & $0.46$~m, $0.085$~m\\
$B_0$, $G_*$ & $0.47$~T, $0.2162$~T~m\\
$I_p$ & $2971$~A\\
$\beta_V$, $p_b$ & $7.86\times10^{-5}$, $0$~Pa\\
$E_k$, $f_{\mathrm{RE}}$ & $2$~keV, $0.20$\\
$\alpha_w$, $\alpha_{\mathrm{eff}}$ & $1.8$, $1.610$\\
$I_{\mathrm{RE}}$, $I_{\mathrm{th}}$ & $594$~A, $2377$~A\\
$\langle p\rangle_V$ & $6.91$~Pa\\\bottomrule
\end{tabular}
\end{minipage}\hfill
\begin{minipage}[t]{0.47\textwidth}
\centering\small
\caption{Parameters of the nominal mixed equilibrium and its Gaussian reference profile. Values are rounded.}
\label{tab:mixed_state}
\begin{tabular}{@{}ll@{}}\toprule
Quantity & Value\\\midrule
$X_h$ & 0.5241\\
$M_{2,A}$ & $2.8194\times10^{-9}$~Wb$^2$\\
$a_1$ & $-0.0015326$\\
$a_2$ & $0.017559$\\
$N_0$ & $9.2173\times10^{15}$~m$^{-2}$\\
$A_a/e$ & $-2.63915\times10^{-4}$~Wb\\
$\zeta_c$ & $-2.26224\times10^{-4}$~Wb\\
$d$ & $3.7690\times10^{-5}$~Wb\\
$s$ & $3.3156\times10^{-5}$~Wb\\
$2s^2$ & $2.1986\times10^{-9}$~Wb$^2$\\\bottomrule
\end{tabular}
\end{minipage}
\end{table*}

We adopt $\beta_V\simeq7.86\times10^{-5}$ as a prescribed dimensionless input from the experimental analysis, approximating total thermal pressure by electron pressure and neglecting ion pressure. The thermal-profile choice $\alpha_w=1.8$ gives $\alpha_{\mathrm{eff}}\simeq1.610$, below the sufficient spectral bound $2.4086$ (Appendix~\ref{app:resolvent}).

% ---- begin mixed_state_results.tex ----
The nominal $256\times256$ equilibrium uses $\varepsilon_A=+1$, with the Gaussian peak at $A/e=\zeta_c=A_a/e+d$ (Table~\ref{tab:mixed_state}). The field reference is $A_*/e=0$, $G_*=0.2162$~T~m, and $\psi|_{\partial\Omega}=0$ fixes the gauge.

\begin{figure*}[tbp]
\centering\includegraphics[width=\textwidth]{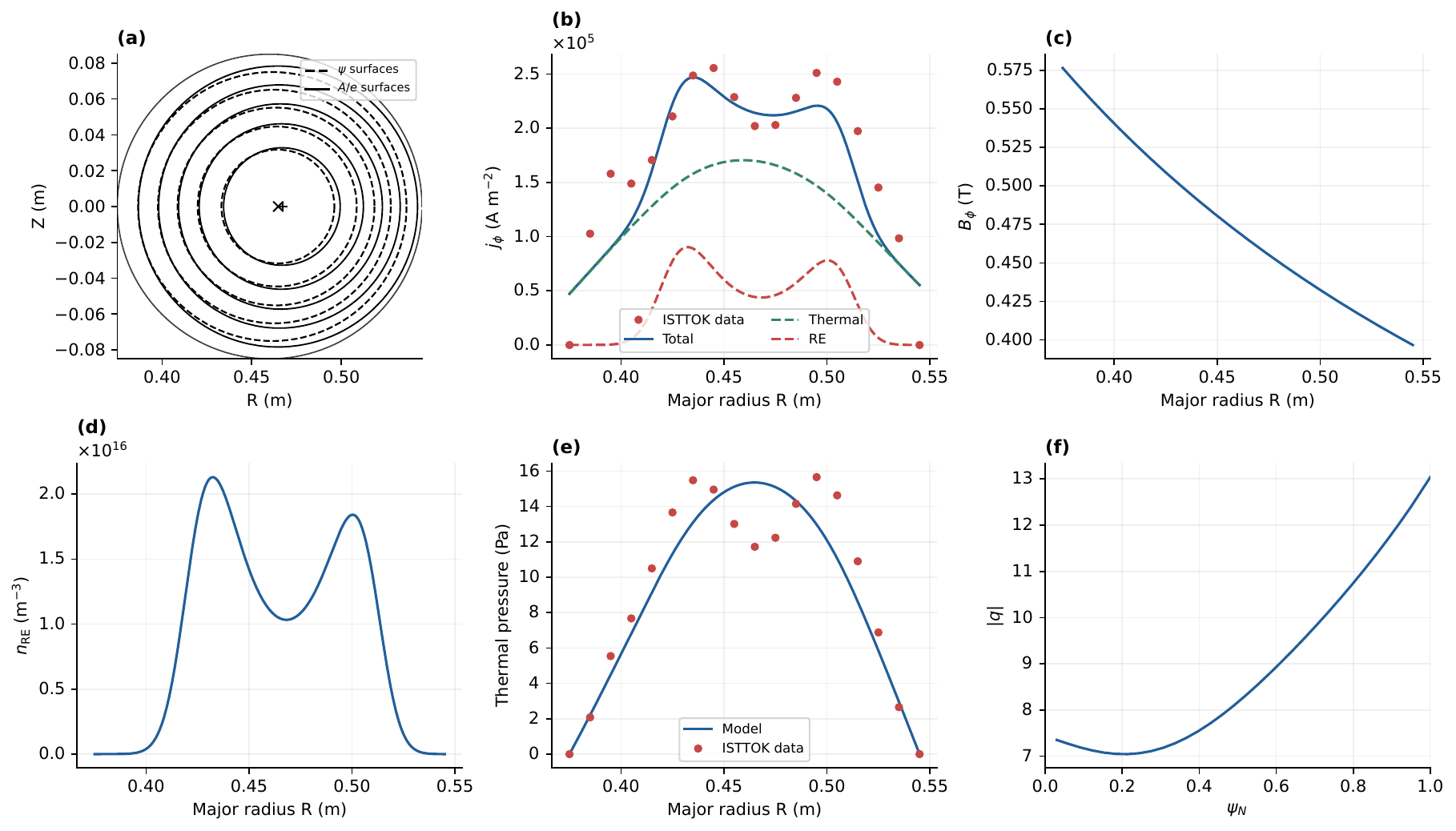}
\caption{Nominal $256\times256$ mixed equilibrium at $f_{\mathrm{RE}}=0.20$ and $E_k=2$~keV. (a) Orbit-label contours $A/e$ (black solid) and magnetic-flux contours $\psi$ (black dashed), with their axes marked. (b) Reconstructed ISTTOK current-density profile normalized to the supplied total plasma current (filled red circles), model total current density (blue solid), thermal current density (green dashed), and RE current density (red dashed). (c) Toroidal magnetic field. (d) Physical energetic-electron density. (e) Total thermal pressure and electron-pressure estimates (filled red circles). (f) Model magnetic safety-factor magnitude. Model profiles in (c)--(f) are blue solid curves, and profiles in (b)--(e) are evaluated at $Z=0$. In (f), $\psi_N=(\psi-\psi_a)/(\psi_b-\psi_a)$ runs from zero at the magnetic axis to one at the boundary, where $\psi_b=0$. Panels (b) and (e) include all 18 derived samples at $t=31.26$~ms, which are tabulated in the supplementary material.}
\label{fig:isttok_example}
\end{figure*}

Figure~\ref{fig:isttok_example} shows a broad thermal current and two energetic-current maxima, near $R=0.4324$ and $0.5003$~m, of $9.03\times10^4$ and $7.80\times10^4$~A~m$^{-2}$. The total current retains two peaks. The density $n_{\mathrm{RE}}=(G/G_B)N_B/R$ reaches $2.13\times10^{16}$~m$^{-3}$, with $1\leq G/G_B\leq1.000178$ on the grid.

Over all 18 samples, the relative discrepancy from the estimated current-density profile is $E_j=0.1791$, or 17.9\%, with root-mean-square discrepancy $3.37\times10^4$~A~m$^{-2}$ and $S=0.3125$. The estimated profile has maxima at $R=0.445$ and $0.495$~m, versus approximately $0.435$ and $0.495$~m in the model. The inboard peak differs by one sample spacing. The model current remains finite at the boundary, where the reconstructed endpoint values are nearly zero. These profile discrepancies are distinct from the equation and constraint residuals in Sec.~\ref{sec:numerical_accuracy}.

The finite thermal edge current in Fig.~\ref{fig:isttok_example}(b) follows from the chosen free functions. At $\psi_b=0$, the pressure vanishes but $p'=-B_0a_1/(\mu_0a^2)>0$ and $F'_b=B_0a_2/G_*>0$, so
\begin{equation}
 j_{\varphi,\mathrm{th},b}
 =-\frac{RB_0a_1}{\mu_0a^2}
 +\frac{G_bB_0a_2}{\mu_0RG_*}>0,
 \label{eq:sec5_thermal_edge}
\end{equation}
where $G_b$ is the boundary trace of $G$. The thermal edge currents are approximately $4.73\times10^4$ and $5.53\times10^4$~A~m$^{-2}$ at the inboard and outboard endpoints. The RE currents there are approximately $0.22$ and $5.8$~A~m$^{-2}$: the Gaussian tail appears zero on the plotted scale but remains positive. Since $F(0)=G_*$, the boundary has $G=G_B$ and $N=N_B$. The density correction does not remove this tail. Neither $p_b=0$ nor $\psi_b=0$ imposes zero current. Allowing the thermal current to vanish at the edge would require different thermal free functions.

The pressure maximum is 15.36~Pa, compared with a largest estimate of 15.66~Pa [Fig.~\ref{fig:isttok_example}(e)]. However, at $R=0.465$~m the model gives 15.36~Pa versus 11.72~Pa. Its single-peaked pressure follows the imposed linear pressure--flux relation on this single-axis equilibrium and cannot reproduce the central depression. This is a free-function restriction, not a resolution effect. After linear interpolation to the 18 sample radii, Eq.~\eqref{eq:sec5_Ej}, with pressure replacing current density, gives a relative discrepancy of 17.7\% and a root-mean-square discrepancy of 1.93~Pa. Only the volume average is constrained through $\beta_V$. The pressure shape was not fitted.

A quadratic midplane extremum fit gives $R_\psi=0.464808$~m. A Cartesian quadratic fit locates the orbit axis at $R_A=0.467198$~m, with $Z_A\simeq0$, hence the 2.39~mm axis separation in Fig.~\ref{fig:isttok_example}(a). Current-profile extrema do not locate the magnetic axis, defined by $\nabla\psi=0$: the model inter-peak minimum at $R\simeq0.4739$~m is about 9~mm farther out. An experimental axis comparison requires poloidal-field or flux reconstruction.

% ---- end mixed_state_results.tex ----

For regular closed magnetic surfaces, the safety-factor \mbox{magnitude} is
\begin{equation}
|q(\psi)|=\frac1{2\pi}\oint_\psi\frac{G}{R|\nabla\psi|}\,dl_p.
\label{eq:sec5_q}
\end{equation}
The local $G=F(\psi)+H(A)$ remains inside the integral. Using the normalized flux defined in Fig.~\ref{fig:isttok_example}(f), the 31 sampled interior surfaces cover $0.03\leq\psi_N\leq0.95$, with $|q|_{\min}\simeq7.05$ and $q_{95}\equiv|q(\psi_N=0.95)|\simeq12.40$. Evaluating the same integral on the plasma boundary gives $|q_{\mathrm{edge}}|\simeq13.04$. For ISTTOK, the cylindrical safety-factor estimate based on the total plasma current is
\begin{equation*}
q^{\mathrm{cyl}}(a)=\frac{2\pi a^2B_0}{\mu_0R_0I_p}\simeq12.42.
\end{equation*}
The model edge value is approximately $5.0\%$ higher than this estimate.

\subsection{Residuals, constraints and field refinement}
\label{sec:numerical_accuracy}

The scalar residual of the converged field, with its final population and thermal coefficients, is
\begin{equation}
r_{\psi,h}=\Delta_h^*\psi+\mu_0R^2p'+GF'-\mu_0evN.
\label{eq:sec5_scalar_residual}
\end{equation}
Writing $b_h=-\mu_0R^2p'-GF'+\mu_0evN$, we normalize it as
\begin{equation*}
\epsilon_\psi=\frac{\|r_{\psi,h}\|_{2,S}}
{\|\Delta_h^*\psi\|_{2,S}+\|b_h\|_{2,S}}.
\end{equation*}
All field norms use the poloidal-area measure,
\begin{equation}
\norm{f}_{2,S}^2=\int_\Omega |f|^2\dd S.
\label{eq:sec5_area_norm}
\end{equation}
With $G=F(\psi)+H(A)$, $H_A=\mu_0vQ(A)$, and $Q=Q(A)$, the vector-equilibrium and particle-conservation conditions hold analytically for any smooth trial flux, since
\begin{equation*}
\nabla G=F^{\prime}\nabla\psi+H_A\nabla A,\qquad
\nabla Q=Q_A\nabla A.
\end{equation*}
With $D_R$, $D_Z$, and $\nabla_h$ denoting numerical derivatives, their differentiation defects are
\begin{align}
r_V&=\frac{\norm{\nabla_hG-F'\nabla_h\psi-H_A\nabla_hA}_{2,S}}
{\norm{\nabla_hG}_{2,S}+\norm{F'\nabla_h\psi+H_A\nabla_hA}_{2,S}},\label{eq:sec5_vector_defect}\\
r_C&=\frac{\norm{(D_RA)(D_ZQ)-(D_ZA)(D_RQ)}_{2,S}}
{\norm{|\nabla_hA||\nabla_hQ|}_{2,S}}.\label{eq:sec5_continuity_defect}
\end{align}
Finite differences do not preserve nonlinear chain rules exactly. Thus $r_V$ and $r_C$ measure differentiation and representation errors, including roundoff, separately from scalar-iteration convergence. The continuum cross product in $r_C$ expresses particle conservation through Eq.~\eqref{eq:sec3_continuity}.
The relative errors in the current and beta constraints are
\begin{equation}
\begin{split}
\epsilon_I&=|I_{\mathrm{th}}+I_{\mathrm{RE}}-I_p|/I_p,\\
\epsilon_{\mathrm{RE}}&=|I_{\mathrm{RE}}-f_{\mathrm{RE}}I_p|/I_p,\\
\epsilon_\beta&=|\beta_V^{\mathrm{calc}}-\beta_V^{\mathrm{input}}|/\beta_V^{\mathrm{input}}.
\end{split}
\label{eq:sec5_constraint_defects}
\end{equation}
An absolute residual is used whenever a residual denominator or target normalization vanishes.

\begin{figure}[!t]
\centering\includegraphics[width=\linewidth]{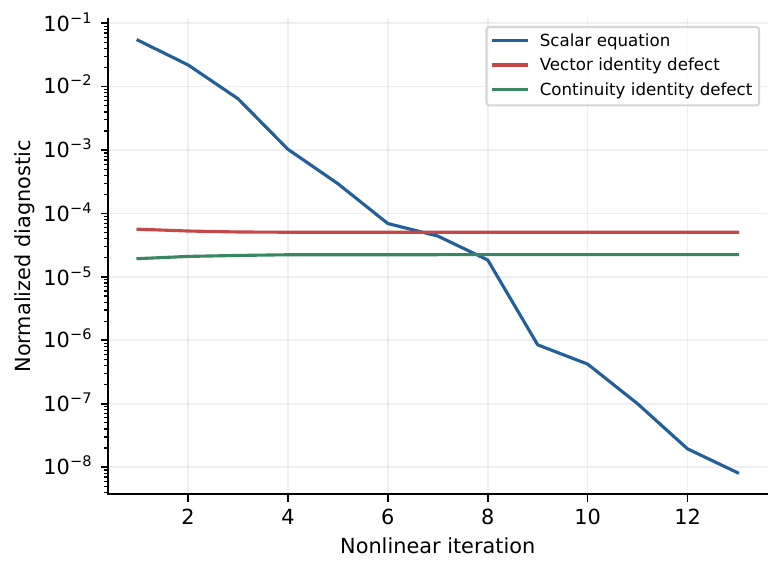}
\caption{Iteration diagnostics for the nominal $256\times256$ mixed equilibrium. The scalar residual $\epsilon_\psi$ measures convergence of the nonlinear field solve. The vector and continuity quantities $r_V$ and $r_C$ measure numerical differentiation defects in identities that the closure satisfies analytically. The algebraic closure residual is at machine precision and is omitted.}
\label{fig:iteration}
\end{figure}

% ---- begin refinement_results.tex ----
For the nominal equilibrium, $\epsilon_\psi=8.18\times10^{-9}$, $r_V=5.05\times10^{-5}$ and $r_C=2.24\times10^{-5}$ (Fig.~\ref{fig:iteration}). The current-constraint errors are $\epsilon_I=6.55\times10^{-13}$ and $\epsilon_{\mathrm{RE}}=5.48\times10^{-14}$. Beta is enforced algebraically to recorded precision. The component-current integral and discrete Amp\`ere current differ by $8.79\times10^{-10}$ relative to $I_p$. The sampled fields have positive radicands and nonnegative pressure and component current densities. The maximum $B_p/B_\phi=0.0173$ supports toroidal-field dominance, independently of the assumed pitch ordering.

At fixed fitted descriptors and physical inputs, refinement from $128^2$ to $256^2$ changes $\psi$, total current density and energetic density by $1.77\times10^{-5}$, $2.88\times10^{-5}$ and $8.51\times10^{-5}$ in relative area-weighted norms. The midplane current converges at approximately second order, and the axis separation changes by $0.492~\mu$m. Appendix~\ref{app:refinement_details} gives the six-grid study, differentiation defects, field comparisons (Fig.~\ref{fig:refinement}) and tighter-tolerance check.
% ---- end refinement_results.tex ----

% ---- end analysis_method.tex ----
% ---- begin isttok_discussion.tex ----
\section{Discussion}
\label{sec:ac_transition_discussion}

\subsection{Coupled fields and current-profile interpretation}

The energetic current modifies both the poloidal field and the constrained thermal coefficients. The thermal-field correction to the reference density, $N/N_B-1=G/G_B-1=(F-G_*)/G_B$, reaches $1.8\times10^{-4}$. This correction enforces consistency within the reduced equations. It measures neither the total toroidal-field change nor the physical approximation error.

At the nominal inputs, we compare $N=(G/G_B)N_B$ with a scalar-equation calculation using $N=N_B$. Both retain the same geometry, physical inputs, thermal-profile parameter, reference label and Gaussian descriptors. Each re-solves $\psi$ and refits $a_1$, $a_2$ and $N_0$ under the same current and beta constraints. Each Gaussian center follows its own orbit axis. Table~\ref{tab:density_correction} gives $D_f=\|f_u-f_c\|_{2,S}/\|f_c\|_{2,S}$ on a $256^2$ grid, with $c$ denoting the compatible solution and $u$ the scalar-only comparison. Both scalar residuals are below $8.1\times10^{-11}$. Repeating the pair at $128^2$ changes each response by less than $0.009\%$, checking the paired discrete response without establishing an absolute continuum-error bound. The scalar-only calculation still violates vector equilibrium and particle conservation.

\begin{table}[tbp]
\centering\small
\caption{Effect of the density correction in the nominal constrained equilibrium. Relative differences use the poloidal-area norm and are dimensionless, not percentages.}
\label{tab:density_correction}
\begin{tabular}{@{}lr@{}}\toprule
Field & Relative response $D_f$\\\midrule
Poloidal flux $\psi$ & $4.65368\times10^{-6}$\\
Total current density $j_\varphi$ & $1.45950\times10^{-5}$\\
RE number density $n_{\mathrm{RE}}$ & $4.73164\times10^{-5}$\\\bottomrule
\end{tabular}
\end{table}

The inboard--outboard asymmetry in Fig.~\ref{fig:isttok_example}(b) is consistent with $n_{\mathrm{RE}}=GQ(A)/R$: at equal $A$, the current-density ratio is $(G_{\mathrm{in}}/G_{\mathrm{out}})(R_{\mathrm{out}}/R_{\mathrm{in}})$. The ratio of the two plotted maxima, about 1.16, is close to the inverse-radius variation for nearly constant $G$.

Distinguishing the thermal and energetic contributions from a total-current profile becomes more difficult when drift displacement is neglected. Formally setting $A=-e\psi$ and defining $\widetilde F(\psi)=F(\psi)+H(-e\psi)$ reduces Eq.~\eqref{eq:sec2_scalar_equilibrium} to
\begin{equation}
 \Delta^*\psi=-\mu_0R^2p'(\psi)
 -\widetilde F(\psi)\widetilde F'(\psi).
\label{eq:sec6_effective_F}
\end{equation}
A flexible thermal field function can therefore absorb the energetic contribution when orbit displacement is omitted. This is a formal comparison: $K=\gamma m_ev$ precludes taking $K\to0$ at finite $v$.

For nearly circular surfaces near the axis, $R_A-R_\psi\simeq q_0\rho_\parallel$, where $q_0$ is the on-axis safety-factor magnitude, $B_{\varphi,a}$ is the on-axis toroidal field, and $\rho_\parallel=|K|/(eB_{\varphi,a})$. The computed 2.39~mm separation at 2~keV and 2.97~kA is about one quarter of the 10~mm sample spacing. Scaling with $|K|$ at fixed field geometry gives a 10~mm separation near 34~keV. Inferring the prescribed $f_{\mathrm{RE}}$ and $E_k$ requires joint sensitivity and uncertainty analysis with independent population measurements. Reported ISTTOK Cherenkov thresholds lie in the tens of keV and above.\cite{Plyusnin2010,Jakubowski2012} The nominal population requires measurements sensitive to 2~keV.

\subsection{Approach to current reversal}

During ramp-down, Eq.~\eqref{eq:sec4_current_pressure_bound} limits pressure support in the co-current sign regime. Relative to poloidal magnetic pressure, the cylindrical estimate is $\beta_p^{\mathrm{cyl}}=2\mu_0\langle p\rangle_V/B_{p,a}^2$ at $p_b=0$, where $B_{p,a}=\mu_0I_p/(2\pi a)$. Its nominal value is about 0.36, compared with the explicit upper bound of about 2.4. If the thermal pressure is held at its $t=31.26$~ms value and $f_{\mathrm{RE}}=0.20$, this equilibrium family is excluded below about 1.1~kA. The excess thermal pressure must vanish with current under these sign assumptions. Applying this restriction to the residual plasma requires its temperature as well as its density.

As the poloidal field weakens, $q_0\rho_\parallel\propto |K|/I_p$ for a fixed current-profile shape. Predicting wall interception requires orbit calculations in the evolving field. At $I_p=0$, $I_{\mathrm{RE}}$ remains defined but $f_{\mathrm{RE}}=I_{\mathrm{RE}}/I_p$ is undefined. A transition model must evolve signed component currents and population moments through the crossing.

The external vertical field becomes especially relevant as the plasma-generated poloidal field weakens. Neglecting the latter and retaining a vertical field $B_v$, Eq.~\eqref{eq:sec2_vRE} gives zero instantaneous vertical drift when $B_v=K/(eR)$. At $R_0=0.46$~m, its magnitude is about 0.33~mT at 2~keV and 2.4~mT at 100~keV, with the sign set by the electron momentum. The vertical-field waveform therefore affects instantaneous drift cancellation, consistent with the orbit analysis of Malaquias \emph{et al.}\cite{Malaquias2022} Confinement also depends on the spatial field, radial motion, and wall geometry.

Once the toroidal electric field reverses, it decelerates the electrons carrying the previous half-cycle's current. With constant opposing loop voltage, fixed radius and no collisions, the stopping time is $\tau_{\mathrm{stop}}=2\pi R_0|K|/(e|V_{\mathrm{loop}}|)$. At $|V_{\mathrm{loop}}|=5$~V it is about $90~\mu$s for 2~keV and $0.65$~ms for 100~keV. These constant-voltage estimates are illustrative. They do not integrate the measured waveform or establish quasistationarity or survival of the prescribed population. Collisional drag adds deceleration while those electrons move faster than the thermal population. Electrons that retain their original direction into the reversed-current phase carry counter-current relative to the new plasma current. For otherwise comparable reversed magnetic geometry, their near-axis drift displacement changes from outward to inward.

Lower momentum reduces orbit displacement but also shortens the stopping time. Survival depends on both effects, collisions and the evolving fields. Tangential hard-x-ray measurements in HT-7 AC discharges suggest energetic electrons moving in both toroidal directions,\cite{Lu2010} which a transition model should allow.

\subsection{Model limitations and experimental uncertainties}

The stationary model assumes a monoenergetic population with narrow pitch and a fixed circular magnetic boundary. It omits collisions, radiation, ionization, particle sources, and particle-wall interactions. Global uniqueness and dynamical stability have not been established. The reference Gaussian is prescribed in orbit-label space, with $f_{\mathrm{RE}}$ and $E_k$ as inputs. In the high-energy pure-RE comparison, $B_p/B_\phi\simeq0.81$ requires reassessment of the approximation that motion is dominated by its toroidal component. These high-energy cases therefore compare solutions of the retained equations without establishing their physical accuracy in that regime. The nominal ISTTOK value is 0.0173. The published shifts used in that comparison are approximate, and their contour-extraction convention is unspecified.

For the nominal state, $(B_p/B_\varphi)^2\simeq2.99\times10^{-4}$ is comparable to the maximum density-factor correction $G/G_B-1\simeq1.78\times10^{-4}$. This ordering comparison is not a calibrated approximation-error estimate. A finite-pitch distribution and uniform bounds on omitted guiding-center terms have not been established. The correction ensures compatibility of the retained equations. Improved physical accuracy at its numerical scale has not been demonstrated.

The refinement results support numerical convergence. Rigorous bounds on the continuum error and on extrema between grid points remain to be established. A nonzero population on boundary-crossing drift contours requires compatible particle fluxes at the boundary. For the nominal state, Appendix~\ref{app:boundary_tail} bounds the population on labels attained at the computational boundary by approximately $0.0264\%$ of the toroidal RE current and $0.0321\%$ of the RE particle content. The label set can also contain disconnected closed components, making the bound conservative. The Gaussian tail remains positive, and these bounds determine neither a normal loss current nor a confinement time. The fixed magnetic boundary therefore does not establish confinement against an absorbing material wall, and the nominal bounds cannot be extrapolated to higher energy or weaker-current states.

The comparison depends on the assumed density distribution, radial mapping and edge treatment in Sec.~\ref{sec:conditional_fitting}. Normalization fixes the integrated current without validating the inferred conductivity shape. Sensitivity to excluding the six peripheral values remains untested. Equal weights do not imply equal or independent uncertainties, and $S$ is not a statistical confidence measure. The restricted thermal profile may influence the fitted RE shape, particularly near the edge. The fit does not independently determine $f_{\mathrm{RE}}$, $E_k$, or a unique thermal--RE decomposition.

\subsection{Further developments}

A dynamical calculation can start from ramp-down equilibria constrained by $I_p(t)$ and thermal pressure, then evolve particle trajectories under the external fields and loop voltage. Near zero current, nested flux surfaces cannot be assumed to persist. Collisions, ionization, wall interception, and coupling to the thermal plasma and circuit would determine whether surviving electrons provide enough ionization and energy for recovery in the next half-cycle. HIBD and energy-sensitive measurements would constrain that comparison.

DREAM offers a fluid--kinetic description of distribution and current evolution,\cite{Hoppe2021DREAM} with zero orbit width and prescribed flux-surface geometry in its published formulation. Application to full AC reversal would require finite orbit widths, orbit losses, and evolving magnetic geometry.

An energy distribution can be represented formally in the equilibrium closure by adding independent constant-energy groups. With signed velocity $v_k$, momentum $K_k=\gamma_km_ev_k$, label $A_k=K_kR-e\psi$ and density $N_k=Rn_k$, separate particle conservation and additive force balance give
\begin{equation}
 \begin{gathered}
 N_k=GQ_k(A_k),\qquad H_k'=\mu_0v_kQ_k,\\
 G=F(\psi)+\sum_k H_k(A_k).
 \end{gathered}
\label{eq:sec6_multienergy}
\end{equation}
All energy groups are coupled through the same $G$ and $\psi$. The solvability of this extension, the choice of functions $Q_k(A_k)$, and a finite-pitch generalization require further analysis.

Coupling between energetic and thermal electrons also matters during start-up once closed flux surfaces form. STREAM calculations show that runaway generation can reduce ohmic heating and hinder burn-through,\cite{Hoppe2022Startup} and cross-machine experiments document start-up runaway production.\cite{deVries2023Startup} Modeling of mildly relativistic electrons and closed-flux-surface formation has also been compared qualitatively with KSTAR start-up.\cite{Lee2026Startup} Plasma recovery thus depends on current and heating as well as surviving electron number.

\section{Conclusions}\label{sec:conclusions}

The axisymmetric model couples a thermal plasma and a directed energetic-electron population through their common magnetic field. Stationary particle conservation and vector integrability give $Rn_{\mathrm{RE}}/G=Q(A)$ on connected orbit contours, and the definition $Q=N_B/G_B$ gives the RE density in the mixed equilibrium through $Rn_{\mathrm{RE}}=(G/G_B)N_B$. Coupling this relation to Wang's thermal free functions retains the displacement between drift and magnetic-flux surfaces and recovers both pure-component limits.

The existence and local-uniqueness results distinguish solvability at fixed thermal coefficients from solvability under prescribed current and pressure constraints. In the co-current regime, the current--pressure bound excludes continuation to zero current at fixed positive excess thermal pressure.

For the ISTTOK example with prescribed $f_{\mathrm{RE}}=0.20$ and $E_k=2$~keV, the numerical solution satisfies the current and beta constraints and retains a two-peaked total-current profile. The magnetic and orbit axes are separated by 2.39~mm. Differences between the computed fields decrease with mesh refinement.

Predicting plasma recovery in the next AC half-cycle requires the energetic distribution and electromagnetic fields to evolve through reversal with collisions, ionization, and wall losses. An extension to energy and pitch distributions, coupled to thermal and circuit evolution, would connect the surviving energetic population to that recovery.

\section*{Supplementary Material}
See the supplementary material for the HIBD-based estimates of the toroidal current density and electron pressure of ISTTOK discharge \#40041 at $t = 31.26$~ms, tabulated at the 18 sample positions shown in Figs.~\ref{fig:isttok_example}(b) and \ref{fig:isttok_example}(e).

\begin{acknowledgments}
M. Keu acknowledges scholarship support from the Indonesia Endowment Fund for Education (LPDP).
\end{acknowledgments}

\section*{Author Declarations}
\subsection*{Conflict of Interest}
The authors have no conflicts to disclose.

\subsection*{Author Contributions}

\textbf{Markus Keu:} Conceptualization (equal), Formal analysis (lead), Investigation (lead), Methodology (lead), Software (lead), Validation (equal), Writing -- original draft (lead), Writing -- review \& editing (equal). \textbf{Matthew Hole:} Conceptualization (equal), Formal analysis (supporting), Methodology (supporting), Supervision (lead), Validation (equal), Writing -- review \& editing (equal). \textbf{Artur Malaquias:} Conceptualization (equal), Data curation (lead), Methodology (supporting), Supervision (supporting), Validation (equal), Writing -- review \& editing (equal).

\section*{Data Availability Statement}
The data that support the findings of this study are available within the article and its supplementary material.
% ---- end isttok_discussion.tex ----
% ---- begin appendices.tex ----
\appendix
% Reduce only the space before the first appendix heading.
\vspace{-\baselineskip}
\section{Population closure and reference profile}\label{app:limits}
\subsection{Density compatibility and profile prescriptions}\label{app:density_obstruction}\label{app:source_weightings}
Suppose $N=\mathcal N(A)$, with $\mathcal N>0$ on an occupied regular contour segment. The quotient rule gives
\begin{equation}
 \nabla\!\left(\frac{\mathcal N(A)}{G}\right)
 =\frac{\mathcal N_A}{G}\nabla A
  -\frac{\mathcal N}{G^2}\nabla G.
 \label{eq:app_quotient_gradient}
\end{equation}
Crossing with $\nabla A$ eliminates only the first term:
\begin{equation}
 \begin{split}
 0&=\nabla\!\left(\frac{\mathcal N(A)}{G}\right)\times\nabla A\\
  &=-\frac{\mathcal N}{G^2}\nabla G\times\nabla A.
 \end{split}
 \label{eq:app_quotient_cross}
\end{equation}
Thus $\nabla G\times\nabla A=0$ wherever $\mathcal N>0$: $G$ must be constant along that drift contour. Only the component of $\nabla G$ tangent to the contour must vanish. Crossing Eq.~\eqref{eq:sec2_vector_equilibrium} with $\nabla A$ gives
\begin{equation}
 \begin{split}
 0=\nabla G\times\nabla A
   &=F'(\psi)\nabla\psi\times\nabla A\\
   &=KF'(\psi)\,\partial_Z\psi\,\bm e_\varphi.
 \end{split}
 \label{eq:app_mixed_obstruction}
\end{equation}
Here the RE term is parallel to $\nabla A$, and $\nabla\psi\times\nabla A=K\nabla\psi\times\nabla R=K\partial_Z\psi\,\bm e_\varphi$ uses $\nabla K=0$. Together with Eq.~\eqref{eq:app_quotient_cross}, this gives the restriction in Eq.~\eqref{eq:sec3_obstruction}.

The incompatibility occurs on occupied regular segments where $K F'\partial_Z\psi\ne0$. Where the density vanishes, Eq.~\eqref{eq:app_quotient_cross} imposes no restriction on $G$. If $F'=0$, including the pure-RE limit, the vector equation permits $G=G(A)$ and $N=\mathcal N(A)$ consistently.

\paragraph*{Pure-RE density prescriptions.}
For $F'=0$, the vector equation makes $G$ constant along regular drift contours, so Eq.~\eqref{eq:sec3_integrability} requires $N$ to be constant there as well. The Constant Radius prescription $N=R_{\mathrm{ref}}n_0g(A)$ satisfies this condition. The Local Radius prescription gives
\begin{equation}
N=Rn_0g(A),\qquad
\nabla N\times\nabla A=n_0g(A)\nabla R\times\nabla A,
\end{equation}
which is nonzero wherever $n_0g(A)>0$ and $\partial_Z\psi\ne0$. The Local Radius calculation in Sec.~\ref{sec:pure_limits} therefore satisfies the scalar equilibrium equation but violates vector compatibility on those contour segments.

\subsection{Thermal force balance}\label{app:thermal_balance}
With $N=GQ(A)$, the velocity in Eq.~\eqref{eq:sec2_vRE} gives the energetic current
\begin{equation}
\bm J_{\RE}=vQ\grad A\times\grad\varphi-evN\grad\varphi.\label{eq:JREapp}
\end{equation}

Subtracting Eq.~\eqref{eq:JREapp} from Amp\`ere's law gives the thermal current. Since $\grad G=F'\grad\psi+H_A\grad A$ and $H_A=\mu_0vQ$, the orbit-dependent poloidal current cancels. Equation~\eqref{eq:sec2_scalar_equilibrium} then gives
\begin{equation}
\bm J_{\thm}=\frac{F'}{\mu_0}\bm B+Rp'\ephi.\label{eq:Jthermal}
\end{equation}
Because $\ephi\times\bm B=\grad\psi/R$, Eq.~\eqref{eq:Jthermal} satisfies the thermal force balance, Eq.~\eqref{eq:sec2_force_balance}. Its toroidal component recovers Eq.~\eqref{eq:sec2_species_currents}, with the full field function $G$ retained.

\subsection{Gaussian reference-profile moments}\label{app:descriptors}
For $k=d/s\geq0$ and $g(\zeta)=\exp[-(\zeta-d)^2/(2s^2)]$, substituting $t=(\zeta-d)/s$ gives
\begin{align}
\int_0^\infty g\dd\zeta&=s\sqrt{2\pi}\Phi(k),\\
\frac{\int_0^\infty\zeta^2g\dd\zeta}{\int_0^\infty g\dd\zeta}
&=s^2[1+k^2+k\phi_N(k)/\Phi(k)].
\end{align}
Using $X_h=e^{-k^2/2}$ in the final relation gives Eq.~\eqref{eq:sec3_shape_inverse}. The term $k\phi_N(k)/\Phi(k)$ accounts for integration over $\zeta\geq0$, so the second moment differs from the full-line value $d^2+s^2$.

\subsection{Population on boundary-accessible labels}\label{app:boundary_tail}
For the nominal Gaussian, let $x=A/e$, $\kappa=K/e<0$ and $R_\pm=R_0\pm a$. Since $\psi_b=0$, every contour intersecting the computational boundary has a label in
\begin{equation}
 W=[\kappa R_+,\kappa R_-],\qquad
 \Omega_W=\{(R,Z)\in\Omega:A/e\in W\}.
 \label{eq:app_boundary_labels}
\end{equation}
The set $\Omega_W$ may also contain disconnected closed components with these labels. The absolute Gaussian peak $\zeta_c=A_a/e+d$ lies below $W$. Therefore, throughout $\Omega_W$,
\begin{equation}
 N_B\leq N_0g_W,\qquad
 g_W=\exp\!\left[-\frac{(\kappa R_+-\zeta_c)^2}{2s^2}\right].
\end{equation}
Both $x$ and $\kappa R$ belong to $W$, so $|\psi|=|\kappa R-x|\leq\psi_W:=2a|\kappa|$. With the nominal $A_*=0$, $a_2>0$, $\Lambda\geq0$ and positive field branches, $G_B\geq G_*$ and
\begin{equation}
 \begin{gathered}
 \frac{G}{G_B}\leq\max\{1,F/G_*\}\leq C_W,\\
 C_W=\left[1+\frac{2B_0a_2\psi_W+\Lambda\psi_W^2}{G_*^2}\right]^{1/2}.
 \end{gathered}
\end{equation}
Thus $N\leq C_WN_0g_W$. The toroidal RE current $I_W$ and particle content $\mathcal P_W$ in this spatial set obey
\begin{equation}
 \begin{split}
 I_W&=e|v|\int_{\Omega_W}\frac{N}{R}\,dS\\
 &\leq2\pi e|v|C_WN_0g_W
 \left(R_0-\sqrt{R_0^2-a^2}\right),
 \end{split}
\end{equation}
\begin{equation}
 \mathcal P_W=2\pi\int_{\Omega_W}N\,dS
 \leq2\pi C_WN_0g_W\pi a^2.
\end{equation}
Since $\mathcal P_{\mathrm{RE}}\geq2\pi R_-I_{\mathrm{RE}}/(e|v|)$, the particle fraction satisfies
\begin{equation}
 \frac{\mathcal P_W}{\mathcal P_{\mathrm{RE}}}
 \leq\frac{e|v|C_WN_0g_W\pi a^2}{R_-I_{\mathrm{RE}}}.
\end{equation}
Using the rounded nominal tables gives $g_W\simeq8.07\times10^{-5}$, $C_W\simeq1.0000071$ and $I_W\lesssim0.157$~A, corresponding to $I_W/I_{\mathrm{RE}}\lesssim0.0264\%$ and $\mathcal P_W/\mathcal P_{\mathrm{RE}}\lesssim0.0321\%$. These are approximate evaluations of conservative spatial bounds, not Gaussian label fractions or wall-loss rates.

\section{Existence, uniqueness, and pressure bounds}\label{app:existence}
\subsection{Positive Dirichlet resolvent}\label{app:resolvent}
For $P=-\Dstar$ with zero trace, we define the weighted inner product
\begin{equation}
(u,h)_w=\int_\Omega\frac{uh}{R}\dd S,
\qquad (Pu,h)_w=\int_\Omega\frac{\grad u\cdot\grad h}{R}\dd S.\label{eq:weight}
\end{equation}
Since $R_-=R_0-a>0$, these weighted norms are equivalent to their ordinary counterparts. The map $Uf=\sqrt R\,f$ is unitary from ordinary $L^2$ to the weighted space and satisfies
\begin{equation}
U^{-1}PU=-\Delta+\frac{3}{4R^2}.\label{eq:unitary}
\end{equation}
The min--max principle gives
\begin{equation}
\frac{j_{0,1}^2}{a^2}+\frac{3}{4R_+^2}
\leq\lambda_1(P)\leq
\frac{j_{0,1}^2}{a^2}+\frac{3}{4R_-^2},\label{eq:eigen}
\end{equation}
where $R_+=R_0+a$ and $j_{0,1}$ is the first zero of $J_0$. For $0\leq\Lambda<\lambda_1(P)$, the inverse of $P-\Lambda$ preserves positivity. For the ISTTOK geometry, the lower bound in Eq.~\eqref{eq:eigen} gives the sufficient spectral-gap condition $\alpha_{\rm eff}<2.4086$. This value includes the toroidal contribution as well as the disk eigenvalue involving $j_{0,1}$.

\subsection{Existence at finite reference-profile amplitude}\label{app:finite_existence}
We fix the reference-profile shape $\chi(A)$, its dimensional center and width, the scale $N_{\rm ref}$, and the reference values. We write $N_B=\eta N_{\rm ref}\chi(A)$ and define
\begin{equation}
\mathcal I(A)=\int_{A_*}^{A}\chi(A')\dd A',
\qquad A_{\max}=KR_-.
\end{equation}
For every nonnegative flux, $A\leq A_{\max}$. On that half-line we assume
\begin{equation}
\begin{gathered}
\mathcal I\leq0,\qquad J_*:=\sup(-\mathcal I)<\infty,\\
M_0:=\sup\chi<\infty,\qquad
M_1:=\sup(A_{\max}-A)\chi(A)<\infty.
\end{gathered}\label{eq:tail}
\end{equation}
For a unit-height Gaussian of angular-momentum width $\sigma_A$ and center $A_c$, choosing $A_*\geq A_{\max}$ ensures $\mathcal I\leq0$ and gives
\begin{equation}
M_0\leq1,\quad J_*\leq\sqrt{2\pi}\sigma_A,\quad
M_1\leq|A_{\max}-A_c|+\sigma_Ae^{-1/2}.
\end{equation}
To bound the nonlinear source in the field equation, we define
\begin{align}
L_F&=\max\{B_0a_2/G_*,\sqrt\Lambda\},\\
h_\eta&=\sqrt{G_*^2+2\mu_0|v|\eta N_{\rm ref}J_*}-G_*,\\
n_\eta&=\eta N_{\rm ref}\left(M_0+\frac{L_FM_1}{eG_*}\right),\\
C_\eta&=L_Fh_\eta+\mu_0e|v|n_\eta.\label{eq:sourcebound}
\end{align}
These bounds are finite for every finite $\eta$. For $\psi\geq0$ and the coefficient signs in Eq.~\eqref{eq:sec4_existence_conditions},
\begin{equation}
(F')^2=\Lambda+\frac{(B_0a_2)^2-\Lambda G_*^2}{F^2}.
\end{equation}
Hence $F\geq G_*$, $0\leq F'\leq L_F$, and $0\leq F-G_*\leq L_F\psi$. The signs $\mathcal I\leq0$ and $v<0$ give $G_B\geq G_*$ and $0\leq H\leq h_\eta$. Because $\psi\leq(A_{\max}-A)/e$, Eq.~\eqref{eq:sec3_actual_density} yields
\begin{equation}
0\leq N\leq\eta N_{\rm ref}\left[M_0+
\frac{L_F}{eG_*}(A_{\max}-A)\chi(A)\right]\leq n_\eta.
\end{equation}
The nonlinear energetic contribution
\begin{equation}
\mathcal T_\eta(R,\psi)=HF'+\mu_0e|v|N
\end{equation}
obeys $0\leq \mathcal T_\eta\leq C_\eta$ uniformly for nonnegative $\psi$. This bounds both the direct energetic-current term and the coupling term $HF'$.

\begin{theorem}[Existence at fixed thermal coefficients]\label{thm:global}
Let the geometry, signed momentum, thermal coefficients, reference-profile shape and its dimensional parameters, and reference values be fixed. Assume Eq.~\eqref{eq:sec4_existence_conditions}, a nonnegative $\chi\in C^{2,1}_{\rm loc}(\mathbb R)$ satisfying Eq.~\eqref{eq:tail}, and $N_{\rm ref}>0$. For every finite $\eta\geq0$ there is a classical zero-Dirichlet solution of Eq.~\eqref{eq:sec3_mixed_GSH} satisfying
\begin{equation}
\psi_0\leq\psi_\eta\leq\psi_0+C_\eta b_\Lambda,\label{eq:barriers}
\end{equation}
where
\begin{equation}
\begin{gathered}
\psi_0=(P-\Lambda)^{-1}f_{\thm},\quad
b_\Lambda=(P-\Lambda)^{-1}1,\\
f_{\thm}=B_0[a_2-a_1(R/a)^2].
\end{gathered}
\end{equation}
The positive field branches are preserved and $N\geq0$. The thermal and total toroidal current densities are positive in the interior. If $a_1<0$, the thermal-pressure excess is positive there. If the reference profile is positive on an open interval of labels attained by the thermal solution, the energetic current is nonzero for $\eta>0$.
\end{theorem}
\begin{proof}
On the closed convex order interval between the two barriers in $C^0(\bar\Omega)$, we define
\begin{equation}
T_\eta u=(P-\Lambda)^{-1}[f_{\thm}+\mathcal T_\eta(R,u)].
\end{equation}
The source bound and positivity of the resolvent make this a self-map, and the regularity assumptions ensure continuity. Bounded forcing gives a bounded $W^{2,p}$ image for finite $p>2$, relatively compact in $C^0$ on the two-dimensional disk. Schauder's theorem gives a fixed point.\cite{Deimling1985} Elliptic regularity then gives a classical solution.\cite{Gilbarg2001} Local regularity of $\chi$ suffices because the solution attains only a bounded interval of labels. The boundary condition constrains only the magnetic flux.

The positive thermal source gives $\psi_\eta\geq\psi_0>0$ in the interior. The bounds above and Eq.~\eqref{eq:sec2_species_currents} give the field and current signs. If the reference profile is positive at a label between the thermal minimum and $A_{\max}$, that label is also attained by $\psi_\eta$: its minimum label cannot increase because $\psi_\eta\geq\psi_0$, while the boundary maximum stays $A_{\max}$. Continuity then gives an open region with positive energetic density and a positive energetic-current integral.
\end{proof}
\subsection{Local uniqueness under the current and beta constraints}\label{app:local}
Let $X=\{\psi\in C^{2,\vartheta}(\bar\Omega):\psi|_{\partial\Omega}=0\}$ and $Y=C^{0,\vartheta}(\bar\Omega)$, $0<\vartheta<1$. For a specified smooth rule defining the reference profile, including any dependence of its center on the field, we define
\begin{equation}
\mathfrak F(z,f)=
\begin{pmatrix}
\mathcal R(\psi,a_1,a_2,\eta;\Lambda(f))\\
I(\psi)-I_*\\
\mathcal B(\psi,a_1)-\beta_*\\
I_{\RE}(z)-fI_*
\end{pmatrix},\qquad z=(\psi,a_1,a_2,\eta),\label{eq:Fcal}
\end{equation}
where $\mathcal R$ is the left-minus-right residual of Eq.~\eqref{eq:sec3_mixed_GSH},
$I(\psi)=-\mu_0^{-1}\int_\Omega\Dstar\psi\dd S/R$, and $\mathcal B$ is Eq.~\eqref{eq:sec2_global_constraints}. At a thermal solution with $\eta=0$, let
\begin{equation}
L_0=\Dstar+\Lambda(0),\quad P_a=D_{(a_1,a_2)}\mathcal R,
\quad \mathsf B=D_\psi(I,\mathcal B).
\end{equation}
Writing $D_{ca}=D_{(a_1,a_2)}(I,\mathcal B)$ gives
\begin{equation}
D_z\mathfrak F=
\begin{pmatrix}L_0&P_a&s_0\\ \mathsf B&D_{ca}&0\\0&0&c_\eta\end{pmatrix},
\quad \mathsf K_{\thm}=D_{ca}-\mathsf B L_0^{-1}P_a,\label{eq:schur}
\end{equation}
where
\begin{equation}
c_\eta=-evN_{\rm ref}\int_\Omega\frac{F_0\chi(A_0)}{RG_*}\dd S>0\label{eq:ceta}
\end{equation}
when the reference profile is positive on a set of orbit labels attained over nonzero area in the thermal solution. Derivatives of $I_{\RE}$ with respect to other state variables vanish at $\eta=0$. The field response $s_0=D_\eta\mathcal R$ includes the energetic-current and toroidal-field contributions.

\begin{theorem}[Locally unique constrained branch]\label{thm:local}
Suppose $I_*>0$, $\mathfrak F(z_0,0)=0$, the pressure and reference-profile prescriptions are $C^1$, and the field radicands and $G$ are uniformly positive. If $L_0:X\to Y$ is an isomorphism, $c_\eta>0$, and $\det \mathsf K_{\thm}\ne0$, there is a unique nearby $C^1$ branch $z(f)$ satisfying Eq.~\eqref{eq:Fcal}. For sufficiently small positive $f$, $\eta(f)>0$, $I=I_*$, $\mathcal B=\beta_*$, and $I_{\RE}=fI_*$. Strict admissibility conditions that are open in the chosen topology persist locally.
\end{theorem}
\begin{proof}
We solve the last linearized row for the amplitude perturbation, eliminate the field perturbation using $L_0^{-1}$, and solve the two scalar constraint rows with $\mathsf K_{\thm}^{-1}$. These steps give a bounded inverse of Eq.~\eqref{eq:schur}. To apply the implicit-function theorem, we extend the amplitude locally to negative values while preserving radicand positivity. Differentiating the last constraint at the thermal equilibrium gives $\eta'(0)=I_*/c_\eta>0$, selecting the physical branch with positive amplitude.
\end{proof}
For a reference center defined by an orbit extremum, the smooth-reference hypothesis includes differentiability of that extremum with respect to the field. A unique interior extremum with nonsingular Hessian is a sufficient local condition for a $C^1$ axis label, provided the selection does not switch to another extremum. At a stationary point, differentiation of the axis label has no first-order contribution from motion of its location. Degenerate or competing extrema can invalidate this smoothness assumption. The constraint matrix $\mathsf K_{\thm}$ must be checked independently of the field spectral gap.

\subsection{Thermal feasibility at fixed current}\label{app:thermal_feasibility}
For fixed $0\leq\Lambda<\lambda_1(P)$, let $L_\Lambda=\Dstar+\Lambda$. The thermal field is linear in $a_1$ and $a_2$ and can be expressed using the solutions of
\begin{equation}
L_\Lambda\psi_1=B_0(R/a)^2,\qquad L_\Lambda\psi_2=-B_0,
\qquad \psi_i|_{\partial\Omega}=0.
\end{equation}
In the positive-resolvent sector, $\psi_1<0<\psi_2$ and $\psi_0=a_1\psi_1+a_2\psi_2$. We define
\begin{equation}
\begin{gathered}
J_1=-I(\psi_1)>0,\quad J_2=I(\psi_2)>0,\\
U_1=-\mean{\psi_1}/(B_0a^2)>0,\quad
U_2=\mean{\psi_2}/(B_0a^2)>0,\\
c_1=U_1/J_1^2,\quad c_2=U_2/(J_1J_2),\quad
t=-J_1a_1/I_p\in[0,1].
\end{gathered}
\end{equation}
Substituting this expansion into thermal beta at fixed current gives
\begin{equation}
\frac{\beta_V-\beta_b}{I_p^2}
=2[c_2t+(c_1-c_2)t^2].\label{eq:thermalresponse}
\end{equation}
Maximizing the quadratic over this coefficient sector gives the sharp bound
\begin{equation}
0\leq\beta_V-\beta_b\leq C_{\thm}I_p^2,\qquad
C_{\thm}=\begin{cases}
2c_1,&c_2\leq2c_1,\\
c_2^2/[2(c_2-c_1)],&c_2>2c_1.
\end{cases}
\end{equation}
A vanishing derivative of Eq.~\eqref{eq:thermalresponse} signals a singular constraint response, distinct from a field resonance. The bound is restricted to $a_1\leq0\leq a_2$ and the specified pressure reference.

\subsection{Mixed current--pressure bound}\label{app:pressure}
Let $c_p=p'\geq0$, $F'\geq0$, $G>0$, and $j_{\phi,\RE}\geq0$. Let $b_R$ solve the Dirichlet problem below, and let $C_R$ denote the geometric integral
\begin{equation}
Pb_R=R^2,\quad b_R|_{\partial\Omega}=0,\qquad
C_R=\int_\Omega R\dd S=\pi R_0a^2.
\end{equation}
The thermal-current formula gives $c_p\leq I_{\thm}/C_R$. Weighted self-adjointness and $P\psi=\mu_0Rj_\phi$ yield
\begin{equation}
\int_\Omega R\psi\dd S
=\mu_0\int_\Omega j_\phi b_R\dd S
\leq\mu_0\norm{b_R}_\infty I_p.
\end{equation}
Multiplying by $2\mu_0c_p/(B_0^2C_R)$ proves Eq.~\eqref{eq:sec4_current_pressure_bound} with
\begin{equation}
C_{\rm mix}=\frac{2\mu_0^2\norm{b_R}_\infty}{B_0^2C_R^2}.
\end{equation}
To bound this coefficient explicitly, we use the disk barrier $b_\Omega=a^2-(R-R_0)^2-Z^2$, which satisfies $Pb_\Omega=2+2R_0/R$. Comparison then gives
\begin{equation}
C_{\rm mix}\leq
\frac{2\mu_0^2R_+^2}
{(2+2R_0/R_+)B_0^2\pi^2R_0^2a^2}.\label{eq:Cmix}
\end{equation}
The sign of $GF'$ is essential here: a negative field contribution could offset the pressure-driven current even when the integrated thermal current is positive.

\section{Numerical methods and comparisons}\label{app:numerical_details}
\subsection{Uniqueness of the current normalization}\label{app:amplitude}
We write $N_B=N_0\chi(A)$ with $\psi$, $F$, and the profile $\chi$ fixed. Under the sign conditions of Appendix~\ref{app:finite_existence},
\begin{equation}
b(A)=2\mu_0v\int_{A_*}^{A}\chi(A')\dd A'\geq0,
\qquad G_B=\sqrt{G_*^2+b(A)N_0}.
\end{equation}
For $\Delta F=F-G_*\geq0$,
\begin{equation}
\frac{\partial}{\partial N_0}\left[N_0\left(1+\frac{\Delta F}{G_B}\right)\right]
=1+\frac{\Delta F\,[G_*^2+bN_0/2]}{(G_*^2+bN_0)^{3/2}}>0.
\end{equation}
After multiplication by $(-ev)\chi/R$ and integration, Eq.~\eqref{eq:sec5_current_norm} has a strictly increasing left side whenever $\chi>0$ on a set of positive area. It starts at zero and grows at least linearly in $N_0$, so it has exactly one nonnegative root for any nonnegative target current.

\subsection{Pure-thermal comparison}\label{app:thermal_comparison_details}
We evaluate Wang's expansion using his Eqs.~(6)--(16), with the asymptotic $x$-weighted projection and modal norm in Eq.~(16a).\cite{Wang2004} The symmetric Hausdorff distance is the larger of the two directed maximum nearest-point distances between contours in the $(R,Z)$ plane.

Current and beta integrals use a common $256\times256$ cell-centered grid.

Against an independent full-operator circular Galerkin solution, the relative flux errors on $128^2$, $256^2$ and $512^2$ grids are $3.56\times10^{-4}$, $8.91\times10^{-5}$ and $2.23\times10^{-5}$, consistent with second-order convergence. Increasing Wang's expansion from 28 radial modes and $m\leq10$ to 40 modes and $m\leq12$ changes its flux by $1.23\times10^{-5}$ relatively. Doubling the grid from $256^2$ to $512^2$ changes the maximum contour distance by $0.090$~mm. With higher-order volume quadrature, Wang's retained expansion gives $\beta_V=0.0101253$, compared with his value of $0.01$. The independently converged full-operator value is $0.00956775$.

% ---- begin refinement_details.tex ----
\subsection{Iteration tolerances and spatial refinement}\label{app:refinement_details}
\begin{figure*}[tbp]
\centering\includegraphics[width=0.82\textwidth]{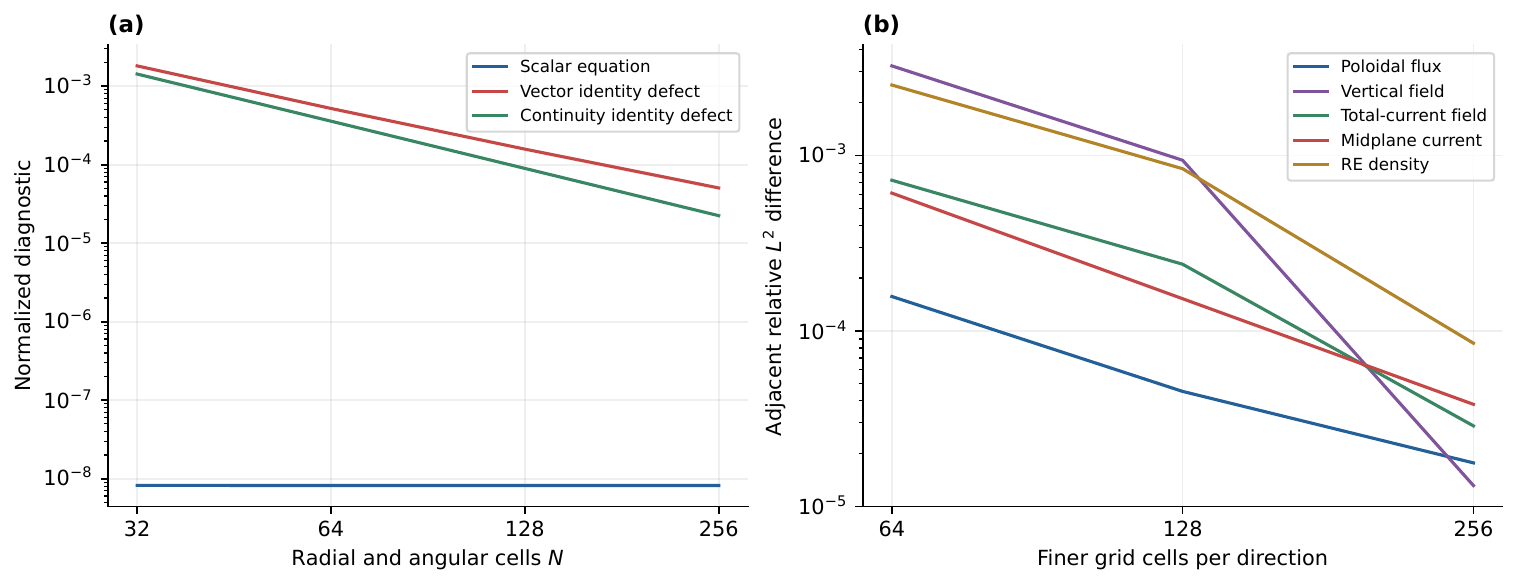}
\caption{Spatial refinement at fixed physical inputs and reference shape. (a) Scalar field residual $\epsilon_\psi$ and vector and continuity differentiation defects $r_V$ and $r_C$ on square grids. The differentiation defects decrease with refinement, while $\epsilon_\psi$ is controlled by the iteration tolerance. (b) Relative differences of independently solved fields and the midplane current on common physical points. Each value compares the indicated grid with the preceding square grid.}
\label{fig:refinement}
\end{figure*}

\begin{figure*}[!t]
\centering\includegraphics[width=0.76\textwidth]{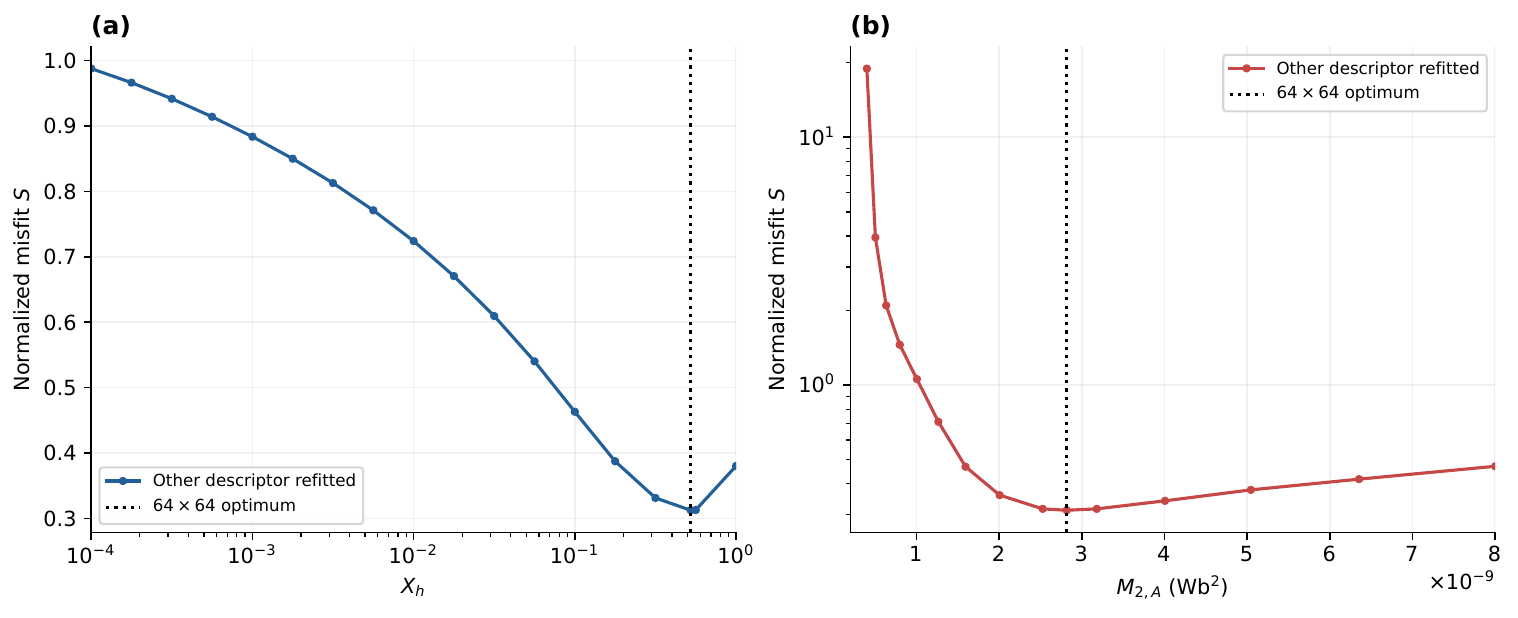}
\caption{Profiled misfit curves at $f_{\mathrm{RE}}=0.20$ and $E_k=2$~keV on a $64^2$ grid. The curve in $X_h$ estimates the minimum of $S$ over $M_{2,A}$ at each fixed $X_h$, and the curve in $M_{2,A}$ estimates the minimum over $X_h$ at each fixed $M_{2,A}$. The equilibrium constraints are enforced at every trial. Vertical lines mark the minimum found in the two-parameter search.}
\label{fig:conditional_profiles}
\end{figure*}

The stopping tolerances are $5\times10^{-8}$ for relative field change, $2\times10^{-7}$ for the scalar residual, $5\times10^{-9}$ for current normalization, and $10^{-12}$ for the algebraic closure residual, with at most 1000 iterations.

For the nominal state, $\min F^2\simeq0.04674$~T$^2$~m$^2$. Since $A\leq A_*=0$ and $v<0$, $G_B^2\geq G_*^2\simeq0.04674$~T$^2$~m$^2$. At grid points and the boundary, $G\geq0.2162$~T~m and $p\geq0$. The smallest thermal current density is about $4.73\times10^4$~A~m$^{-2}$.

To assess spatial convergence, we solve the constrained problem on $32^2$, $64^2$, $128^2$ and $256^2$ grids, together with $128\times256$ and $256\times128$ grids, at fixed $X_h$, $M_{2,A}$ and physical inputs. Every mesh uses the same reference shape, origin treatment, and boundary values. The differentiation defects decrease under joint radial and angular refinement (Table~\ref{tab:refinement}), while the scalar residual remains close to its iteration stopping level.

For direct field comparisons, periodic bilinear interpolation maps each two-dimensional field to a common $128\times128$ polar mesh. Parity at the origin and prescribed values at the computational boundary extend the interpolation to neighboring cells. The comparison uses interior cell centers, excluding the exact origin and boundary. Relative differences are measured in the area-weighted $L^2$ norm and normalized by the finer-grid field. Between $128^2$ and $256^2$, the relative differences are $1.77\times10^{-5}$ in $\psi$, $2.88\times10^{-5}$ in total current density, $8.51\times10^{-5}$ in energetic density, and $1.32\times10^{-5}$ in $B_Z$. Doubling only the angular resolution at $N_r=256$ changes the flux by $4.56\times10^{-6}$, and doubling only the radial resolution at $N_\theta=256$ changes it by $1.48\times10^{-5}$.

\begin{table}[!t]
\centering\small
\caption{Fixed-shape numerical refinement. The quantities $r_V$ and $r_C$ are differentiation defects of analytical identities. The axis separation uses the same magnetic and orbit-axis definitions on every grid.}
\label{tab:refinement}
\begin{tabular}{@{}lrrr@{}}\toprule
$N_r\times N_\theta$&$r_V$&$r_C$&$R_A-R_\psi$ (mm)\\\midrule
$32\times32$&$1.812\times10^{-3}$&$1.427\times10^{-3}$&2.3771\\
$64\times64$&$5.199\times10^{-4}$&$3.581\times10^{-4}$&2.3869\\
$128\times128$&$1.578\times10^{-4}$&$8.960\times10^{-5}$&2.3899\\
$128\times256$&$1.577\times10^{-4}$&$9.444\times10^{-5}$&2.3901\\
$256\times128$&$5.077\times10^{-5}$&$3.844\times10^{-5}$&2.3903\\
$256\times256$&$5.053\times10^{-5}$&$2.241\times10^{-5}$&2.3904\\\bottomrule
\end{tabular}
\end{table}

Comparing the midplane current at 801 common points gives a $128\to256$ relative difference of $3.82\times10^{-5}$, or $0.00382$\%, and an observed convergence order of about 2 over $64$, $128$ and $256$ cells per direction. The axis-separation change is $0.492$~$\mu$m. Figure~\ref{fig:refinement} compares the residuals with these field differences. On the same grid, tighter iteration tolerances reduce the scalar residual to $8.91\times10^{-10}$. The corresponding relative changes are $4.28\times10^{-9}$ in flux and $7.17\times10^{-9}$ in total current density, well below the differences between grids. The two-dimensional field comparisons include interpolation as well as discretization errors.

% ---- end refinement_details.tex ----

\subsection{Reference-profile fitting and sensitivity}\label{app:search_details}

We parameterize the search domain in Sec.~\ref{sec:conditional_fitting} by
\begin{equation}
 u_1=\frac{-\ln X_h}{\ln10^4},\qquad
 u_2=\frac{\ln[M_{2,A}/(2\times10^{-10}\,{\rm Wb}^2)]}{\ln40},
\end{equation}
both in $[0,1]$. On a $64\times64$ field grid, the search uses 256 Sobol points, nine points on the search boundary and 32 additional sampling points, followed by bounded trust-region least-squares minimization from six initial points. The least-squares stopping limits are \texttt{ftol}$=10^{-10}$ and \texttt{xtol}$=$\texttt{gtol}$=10^{-7}$, with at most 60 function evaluations per local start. The coarse-grid field-change and scalar-residual limits are $5\times10^{-7}$ and $2\times10^{-6}$, respectively. The fine-grid stopping tolerances are given in Appendix~\ref{app:refinement_details}.

For the normalized residuals $r_i=(\widehat j_i-j^{\mathrm{rec}}_{\phi,i})/s_j$, we define $J_{ij}=\partial r_i/\partial u_j$. The Jacobian is evaluated from equilibria recomputed under the current and beta constraints. The local minimization terminates when its step-size criterion is met. A separate finite-difference estimate gives a projected-gradient norm of $1.03\times10^{-4}$. The Jacobian singular values are $4.66$ and $1.86$, giving condition number $2.50$. Halving the difference step from $0.005$ to $0.0025$ changes the Jacobian by $3.42\times10^{-4}$ relatively.

For each profiled misfit curve in Fig.~\ref{fig:conditional_profiles}, one descriptor is fixed and the other is sampled at 25 points before a bounded minimization.

For fixed $X_h$ or $M_{2,A}$, the profiled objective is
\begin{equation}
S_X(x)=\min_{M_{2,A}}S(x,M_{2,A}),\qquad
S_M(m)=\min_{X_h}S(X_h,m).
\label{eq:sec5_shape_profiles}
\end{equation}
The minima are approximated over admissible equilibria within the search bounds.

% ---- end appendices.tex ----
% AIP reprint style suppresses this heading by default.
\renewcommand{\refname}{References}
% Create the reference section using BibTeX (REVTeX selects aipnum4-1 automatically):
\bibliography{references}

\end{document}